%% file: main.tex
\documentclass{JHEP3}
\pdfoutput=1
\usepackage[dvips]{graphicx}
\usepackage{latexsym,amsmath,amsfonts,amssymb}
\usepackage{mathenv,mathrsfs,slashed}
\usepackage{multirow}
\usepackage{array}
\usepackage{ulem}
\usepackage{cite}
\usepackage{subfigure}
\usepackage{color}

\usepackage{afterpage}

\renewcommand{\eqref}[1]{Eq.~(\ref{#1})}

\newcommand{\figref}[1]{Figure~\ref{#1}}

\newcommand{\eg}{e.\,g.\/}
\newcommand{\ie}{i.\,e.\/}

\newcommand{\GeV}{\text{GeV}}
\newcommand{\TeV}{\text{TeV}}

\def \cm{c.\,m.}

\newcommand{\mll}{M_{\ell\ell}}
\newcommand{\MT}{M_T (\ell\nu)}

\newcommand{\met}{\slash \!\!\!\!E_{T}}
\newcommand{\ptmiss}{\slash \!\!\!p_{T}}
\newcommand{\jet}{\text{jet}}
\newcommand{\lep}{\ell}

\def\pT{$p_T$\/}
\def\Zy{$Z/\gamma^*$\/}
\def\DY{Drell--Yan\ }
\def\Zjet{$Z/\gamma^*$+jet\ }
\def\Wjet{$W$+jet\ }
\def\DYjet{Drell--Yan+jet\ }

\def \msbar{\overline{\text{MS}}}

\def \qqbar{$q\bar{q}$}

\newcommand{\squark}{\tilde{q}}

\newcommand{\gluino}{\tilde{g}}
\newcommand\msquark{m_{\tilde{q}}}
\newcommand\mgluino{m_{\tilde{g}}}

\newcommand\plot[2]{}

\title{SUSY QCD corrections to electroweak gauge boson production with an associated jet at the LHC}

\author{Ryan Gavin\\
 Phenomenology Institute, Department of Physics, 
 University of Wisconsin-Madison, 
 1150 University Avenue, 
 Madison, Wisconsin 53706, USA\\
and\\
 Paul Scherrer Institut, CH-5232 Villigen PSI, Switzerland\\
 Email: \email{ryan.gavin@psi.ch}}

\author{Maike K.~Trenkel\\
 Phenomenology Institute, Department of Physics, 
 University of Wisconsin-Madison, 
 1150 University Avenue, 
 Madison, Wisconsin 53706, USA\\
 Email: \email{trenkel@hep.wisc.edu}}

\abstract{%
We study the stability of the neutral- and charged-current \DY process in association with a jet as a standard candle at the LHC
under the inclusion of $\mathcal{O}(\alpha_s)$ supersymmetric QCD (SQCD) corrections within the MSSM.
We include the decay of the electroweak gauge boson into dileptons, \ie\ we consider the production of 
charged lepton--anti-lepton or lepton-neutrino final states with one hard jet.
We find that the SQCD corrections are negligible for the integrated cross section. Only at high lepton transverse momentum can they induce effects of the percent level.
\\[2ex]

}


\preprint{MADPH-11-1575}

\begin{document}

\input{Section1_Introduction}

\input{Section2_Calculation}

\input{Section4_NumResults}
\input{Section4_NumResults_W}

\input{Section5_Conclusions}

\section*{Acknowledgments} 
We are grateful to T.~Hahn, S.~Kallweit, and F.~Petriello for useful discussions and helpful comments.  
This work was supported by the US DOE under contract No.~DE-FG02-95ER40896.

\appendix

\input{Appendix_Renormalization}

\bibliographystyle{JHEP}
\bibliography{references}

\end{document}

%% file: Section1_Introduction.tex

\section{Introduction}

Electroweak gauge boson production with subsequent leptonic decays plays a crucial role in the physics studies performed at hadron colliders.  
With large cross sections and a clear collider signature, the \DY processes have become a standard candle at the LHC and many precise measurements are currently being performed \cite{Collaboration:2011cm,Collaboration:2011nx,Collaboration:2011gj,Aad:2010yt}.
Their study can be used to help calibrate detectors and place constraints on parton distribution functions (PDFs), see \eg\ \cite{Ball:2010de,Thorne:2010kj,Lai:2010vv}.  
It has also been proposed that measured electroweak boson production cross sections 
can be used as a luminosity monitor~\cite{Dittmar:1997md}.  
\DY processes also constitute an important source of background for searches for new physics, such 
as $Z'$ and $W'$ boson production and other high mass dilepton resonances
\cite{Rizzo:2006nw,Chiang:2011kq,ATLAS-TDR:1999fr,Ball:2007zza}.

Given the large amount of data that the LHC will deliver, measurement errors will eventually be
 dominated by systematics rather than statistics, see \eg \cite{systematics}.  
Consequently, electroweak boson production needs to be known to high precision.  
Much effort has been made in this area.  
Indeed, it was the first hadronic scattering process to be computed
at next-to-next-to-leading (NNLO) in QCD~\cite{Hamberg:1990np, Harlander:2002wh}.
Now also the differential cross sections for \DY processes are known through NNLO in 
QCD~\cite{Anastasiou:2003yy,Anastasiou:2003ds,Melnikov:2006di,Melnikov:2006kv,Catani:2009sm,Catani:2010en}, 
and the NLO QCD corrections have been matched to parton showers~\cite{Alioli:2010qp,Frixione:2007vw,Hamilton:2008pd,Frixione:2006gn}.
Electroweak corrections have been known for sometime~\cite{Baur:1997wa,Baur:1998kt,Baur:2001ze,Dittmaier:2001ay}, 
and continue to be refined~\cite{Dittmaier:2009cr,Brensing:2007qm,Richardson:2010gz,Zykunov:2008zz,Hamilton:2008pd,Balossini:2008cs}.  
Attempts to approximate QCD $\times$ EW corrections to
\DY have also appeared in the literature~\cite{Balossini:2011zz,Balossini:2008cs}
and very recently 
the mixed QCD $\times$ EW two-loop virtual corrections have been calculated~\cite{Kilgore:2011pa}.

At hadron colliders, the high center-of-mass energy (\cm) typically leads to production of electroweak gauge bosons in 
association with QCD radiation. Here, we focus on \DY processes with one additional, hard jet.
The intermediate gauge boson then recoils against the jet and can be strongly boosted. 
If the jet has large transverse momentum, the leptons originating from the gauge boson decay are produced 
at high transverse momentum as well and provide a source for either high-energy dilepton pairs of opposite charge 
(\Zjet production) or high missing transverse energy in combination with a charged lepton (\Wjet production).
Theoretical predictions for the \DYjet cross section include NLO QCD corrections~\cite{Giele:1993dj,vanderBij:1988ac,Campbell:2002tg} and parton shower matching~\cite{Alioli:2010qp}, 
as well as in the electroweak sector the NLO corrections in the on-shell 
approximation~\cite{Kuhn:2004em,Kuhn:2005az,Kuhn:2007qc,Kuhn:2007cv,Hollik:2007sq} 
and the full NLO calculation, including photonic corrections and including the leptonic decay 
of the gauge boson~\cite{Denner:2009gj,Denner:2011vu}.
Also for \DY cross sections with higher jet multiplicities, the NLO QCD corrections are now available 
\cite{Campbell:2002tg,Ellis:2008qc,Berger:2009ep,Berger:2010vm,Berger:2010zx}.

Often, precision measurements of Standard Model (SM) processes 
are used to constrain new physics, not by direct detection, 
but by their influence through quantum fluctuations.  
One may ask the question how radiative corrections from physics beyond the Standard Model (BSM) 
would effect such well-known processes.
To allow for precision tests of the SM and its perturbative expansions,
it is crucial to know if there were large BSM corrections since otherwise information about the underlying physics 
cannot reliably be extracted from the data.
An interesting paradigm for physics beyond the SM is supersymmetry (SUSY).  
Many collider studies have been performed to examine the phenomenology of SUSY production at hadron colliders, 
and the possibility of their detection.  
The investigation of new physics contributions to SM processes via higher-order corrections is a logical next step.
Only in recent years, \DY processes have been studied   
within the framework of the minimal supersymmetric Standard Model (MSSM). 
The NLO supersymmetric QCD (SQCD) and electroweak corrections within the MSSM
to the charged- and neutral-current \DY processes with no final state parton
were calculated in~\cite{Brensing:2007qm,Dittmaier:2009cr} and 
the NLO electroweak corrections for the on-shell \Wjet production process within MSSM have been computed in~\cite{Gounaris:2007gx}.
The impact of SUSY corrections has been found to be small.

In this work, we further complete the one-loop picture and calculate the NLO SQCD corrections within the MSSM 
to \DY processes in association with a hard jet in the final state. 
We consider both \Zjet and \Wjet production at the LHC, including the leptonic decay of the gauge bosons, and 
take all off-shell effects due to the finite widths of the $Z$ and $W$ boson
and all contributions of an intermediate photon into account. Our aim is to show the stability of \DYjet processes 
under the inclusion of SQCD corrections within the MSSM.  

This paper is organized as follows.
The setup of the calculation is detailed in section~\ref{sec_calc}.
The computation of the SQCD corrections to the neutral- and charged-current \DY processes proceed in close analogy 
and are discussed in parallel.  
In section~\ref{sec_numres} we present the numerical results for both \Zjet and \Wjet production at the LHC.
We summarize our findings in section~\ref{sec_conclusion}. 
Details on the renormalization procedure are given in appendix~\ref{app_renorm}.

%% file: Section2_Calculation.tex

\section{Details of the calculation}
\label{sec_calc}

In this section we describe the technical setup of our calculation of SQCD corrections for the
neutral- and charged-current \DY process with an associated hard jet. We always include the decay of the gauge bosons into leptons and take all off-shell effects due to the finite width of the $Z$ or $W$ boson into account. 
We first address the leading order (LO) processes in section \ref{subsec_Z_LO} and
then discuss the NLO calculation in section \ref{subsec_Z_NLO}.
In view of SQCD corrections, the computations for \Zjet and \Wjet production 
are very similar and proceed in close analogy.

\subsection{Leading order processes}
\label{subsec_Z_LO}

At the LHC, the \DY process in association with a hard jet
is initiated by three different partonic processes at LO.
Including the leptonic decay of the intermediate electroweak gauge boson, they 
are in the case of the neutral-current \DY process,
\begin{eqnarray}
&& {q \; \bar{q} \rightarrow   Z/\gamma^* \; g    \rightarrow \ell^-  \ell^+ \; g \label{qqbar}}\; , \\
&& {g \; q \rightarrow     Z/\gamma^* \; q \rightarrow   \ell^- \ell^+ \; q \label{gq}}\; , \\
&& {g \; \bar{q} \rightarrow     Z/\gamma^* \; \bar{q}   \rightarrow \ell^-  \ell^+ \; \bar{q} \label{gqbar}}\;, 
\end{eqnarray}
and for the charged-current \DY process, 
\begin{eqnarray}
&& {q \; \bar{q}' \rightarrow   W \; g    \rightarrow \ell \; \nu_{\ell} \; g \label{qqbarW}}\; , \\
&& {g \; q \rightarrow     W \; q' \rightarrow   \ell \; \nu_{\ell} \; q' \label{gqW}}\; , \\
&& {g \; \bar{q} \rightarrow    W \; \bar{q}'   \rightarrow \ell \; \nu_{\ell} \; \bar{q}' \label{gqbarW}}\; .
\end{eqnarray}
For $W^+$ production, a positively charged lepton is produced together with an anti-neutrino ($\ell^+ \bar{\nu}_{\ell}$),
for $W^-$ production it is a negatively charged lepton and a neutrino ($\ell^- \nu_{\ell}$).

The tree-level Feynman diagrams for the process \eqref{qqbar} can be found in \figref{fig_Feynman_ZjetLO}, those for \Wjet production follow analogously 
by replacing the $Zqq$ ($Z\ell\ell$)  vertex with a $Wqq'$ ($W\ell\nu_{\ell}$) vertex.
Diagrams for the (anti-)quark--gluon processes are obtained by crossing the gluon with an initial-state (anti-)quark.
%

\FIGURE[t]{
\hspace*{2cm}\includegraphics{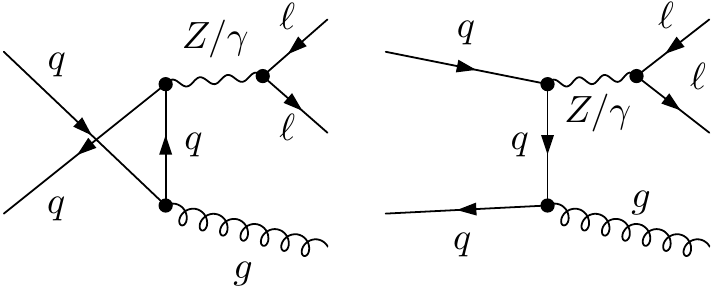}\hspace*{2cm}
\caption{Feynman diagrams for the quark--anti-quark initial state process, \eqref{qqbar}.}
\label{fig_Feynman_ZjetLO}
}

The quarks considered are $q,q' = u,\,c,\,d,\,s$ with $q$ and $q'$ being quarks of opposite isospin. The quarks are treated as massless throughout the calculation.  
The \Zjet LO cross section is quark generation independent, aside from the appropriate PDF inclusion and convolution.  The only up- and down-type quark 
dependence comes from the vector and axial couplings in the $Zq\bar{q}$ vertex.
The \Wjet cross sections do depend on the quark flavor due to the quark mixing parameterized by the CKM matrix. 
However, the CKM quark mixing factorizes from the tree-level matrix elements and, for the inclusive cross section, 
there needs only one squared amplitude to be calculated for each of three partonic channels, weighted by the sum of squares of the respective absolute value elements of the CKM matrix, as well as the corresponding PDFs.

%
Throughout the calculation, we exclude bottom-quark initiated processes 
due to their PDF-suppressed small contribution to the cross section and we  
neglect the mixing between first and third generation quarks.
A fixed width is included in the propagator of the intermediate $Z$ or $W$ boson, giving the expected Breit-Wigner resonance.

Already at LO, there exists the possibility that the parton in the final state may become soft or collinear to the gluon or quark
that it couples to.  If this occurs, the cross section becomes infinite.  
Divergences of this type are regulated by requiring the jet to be hard.  This
is achieved by applying a minimum transverse-momentum ($p_T$) cut on the jet.
Also we treat the leptons as massless throughout the calculation and require a minimum lepton \pT\ as 
well as a minimum lepton-pair invariant mass, $\mll$, in case of \Zjet production.

\subsection{NLO SQCD corrections}
\label{subsec_Z_NLO}

We calculate the one-loop SQCD corrections to the partonic processes in Eqs.~\ref{qqbar}-\ref{gqbar} and Eqs.~\ref{qqbarW}-\ref{gqbarW}, 
which are purely virtual.  
The SUSY particles charged under $SU(3)_C$ are the squarks, $\tilde{q}_i$, and gluinos, $\tilde{g}$, the superpartners 
of the SM chiral quarks and gluons, respectively.  In the SM, left- and right-handed quarks transform differently under
$SU(2)_L$.  
Consequently, under supersymmetric transformations, there are separate superpartners, $\tilde{q}_L$, $\tilde{q}_R$, 
for quarks of different handedness, $q_L$,  $q_R$.    
After electroweak symmetry breaking (EWSB), left- and right-handed squark
eigenstates mix to form squark mass eigenstates, $\tilde{q}_1$,  $\tilde{q}_2$.  
Squark mixing, however, is proportional to the mass of their SM quark partner.   
In this calculation, only squarks of the same flavor as the initial state quarks are present.  
Since the masses of the quarks considered ($u$, $d$, $c$, $s$) are small, we can neglect any left-right 
mixing between squark eigenstates, and the two gauge eigenstates of a given flavor $q$ are also mass eigenstates
that we denote by $\squark_i$, $i=L,R$.

In the calculation of \Zjet production, the relevant interactions between SM and SUSY particles  are the
 $\gamma\tilde{q}_i\tilde{q}_j$, $Z\tilde{q}_i\tilde{q}_j$, 
$g\tilde{q}_i\tilde{q}_i$, $g\tilde{g}\tilde{g}$, and $q\tilde{q}_i\tilde{g}$ vertices.
Although there are diagrams with the $\gamma g\tilde{q}_i\tilde{q}_j$ or $Z g\tilde{q}_i\tilde{q}_j$ 
vertex, their contribution to the virtual corrections is zero, and they have not been explicitly included in this manuscript.
There are 44 diagrams per partonic channel that give
non-zero contributions to the differential cross section (22 for $Z$ boson and photon mediation each).  
They divide into self-energy insertions, gluon- and \Zy-vertex corrections, and box contributions and are organized in \figref{fig_nlodiags}. 
In each diagram, the two squark eigenstates $\squark_i$ of the same quark flavor $q$ as 
the initial state can run in the loop.
The Feynman diagrams for \Wjet production can easily be inferred from
Figure \ref{fig_nlodiags} by again replacing the $Zqq$ ($Z\ell\ell$)
 vertex with a $Wqq'$ ($W\ell\nu_{\ell}$) vertex and  
by replacing the $Z\squark_i \squark_i$ by $W\squark_i\squark'_i$ vertices.
Of course the $W$~boson couples to left-handed particles only, 
and only the left-handed components of the squark eigenstates contribute (\ie\ $\squark_i = \squark_L$ in case of no left-right mixing).

\FIGURE[t]{
(a) Self-energy insertions:\hspace*{10cm}\\[1ex]%
\includegraphics{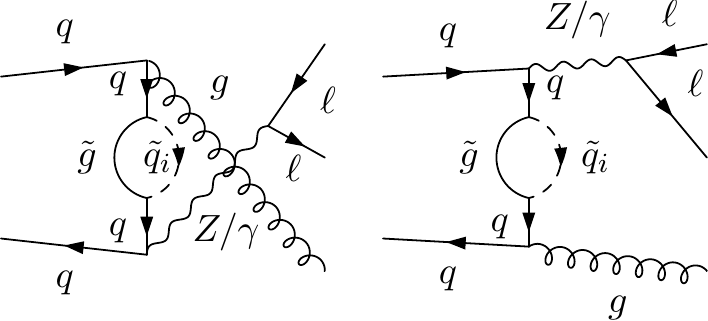}\\%
\\[0ex]
(b) Gluon-vertex corrections:\hspace*{9.5cm}\\[1ex]%
\includegraphics{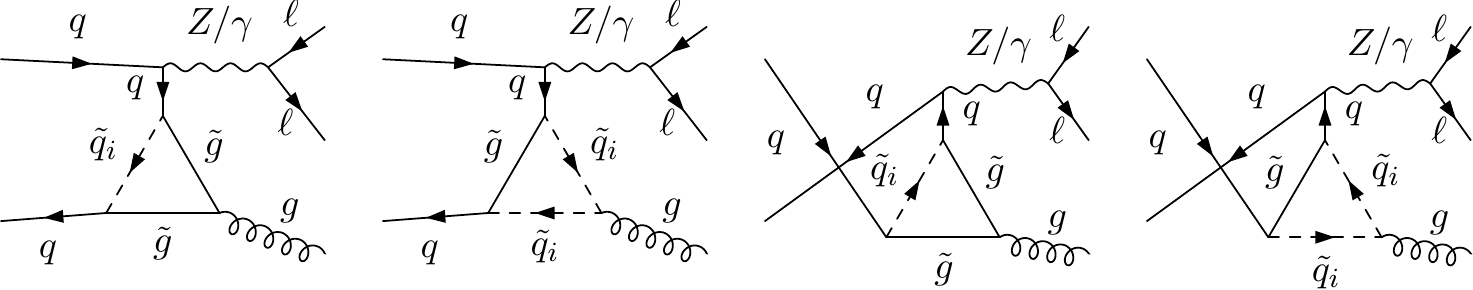}\\%
\\[0ex]
(c) $Z/\gamma^*$-vertex corrections:\hspace*{10cm}\\[1ex]%
\includegraphics{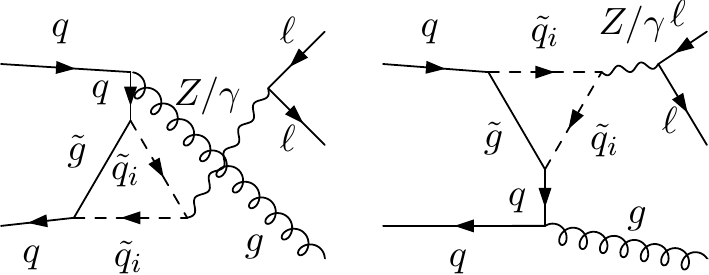}\\%
\\[0ex]
(d) Box contributions:\hspace*{11cm}\\[1ex]%
\includegraphics{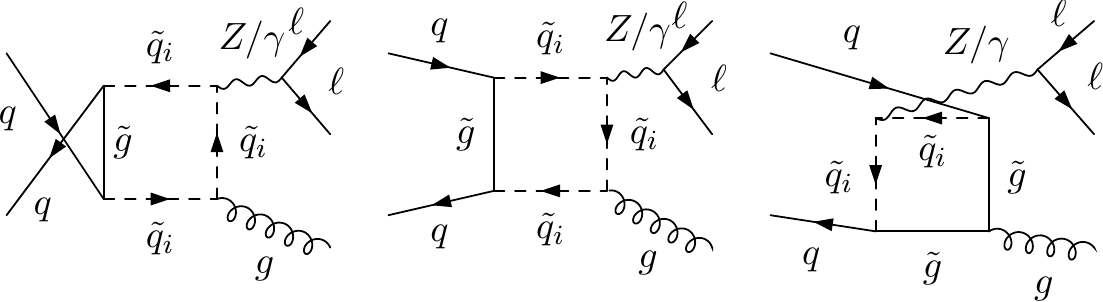}%
\caption{Feynman diagrams for the self-energy, gluon- and \Zy-vertex, and box contributions to 
the partonic process $q\bar{q} \rightarrow \ell^+\ell^- g$, mediated by \Zy\ exchange, at SQCD NLO.
\label{fig_nlodiags}}
}

Due to only massive particles propagating in the loop, the virtual corrections are completely infrared (IR) safe.  Therefore,
there are no real emission processes to be considered to cancel singularities.  
However, as noted above, we require a hard jet in the final state.  
This statement implies a minimum cut has been placed on the transverse momentum of the jet to avoid IR divergences
that would otherwise appear at LO.  

Ultraviolet (UV) divergences arise from the self-energy insertions and the gluon and $Z/\gamma^*$ or $W$ vertex 
corrections, while the box diagrams are IR and UV finite.  
Dimensional regularization is used to regulate the UV divergences, 
where the number of dimensions is $d=4-2\epsilon$ and the singularities then take the form of $1/\epsilon$ poles.
In order to remove the UV divergences one has to include the proper counterterms for SQCD one-loop renormalization.
Our renormalization procedure closely follows ref.~\cite{Berge:2007dz}.
We impose on-shell conditions to fix the renormalization constants in the quark sector and use the 
$\msbar$ scheme, modified to decouple the heavy SUSY particles \cite{Collins:1978wz,Nason:1989zy}, for 
the renormalization of the strong coupling constant and the gluon field. 
The explicit expressions in terms of the renormalization constants can be found in appendix~\ref{app_renorm}.

We set up the calculation in the conventional diagrammatic approach.  
As an important cross-check on the numerical results,
two completely independent computations were performed using different techniques.  
The first calculation is based on the framework of 
FeynArts~3.6 \cite{Hahn:2000kx}, FormCalc~7.0, and LoopTools~2.6 \cite{Hahn:1998yk}.
In the second calculation, Feynman diagrams are generated with QGRAF \cite{Nogueira:1991ex}.  
Algebraic manipulation is performed with an in-house software comprised of Maple and Form \cite{Form}.  
Then AIR \cite{Anastasiou:2004vj} is used for the reduction of tensor loop integrals to scalar ones.  
The numerical integration is performed using Vegas of the Cuba~1.7 \cite{Hahn:2004fe} library, 
as well as the scalar loop integral library \mbox{QCDLoop} \cite{Ellis:2007qk}.

The $Zqq$, $Wqq'$, $Z\tilde{q}_i\tilde{q}_j$, $W\tilde{q}_i\tilde{q}'_j$, and $q\tilde{q}\tilde{g}$ vertices 
contain axial pieces.
Special attention should be paid to the the treatment of $\gamma^5$ when working in $d$ dimensions.  
In the first calculation the naive anticommuting scheme~\cite{Hahn:1998yk,Jegerlehner:2000dz} 
was used.  Here, it is assumed that $\{\gamma^5 , \gamma^\mu \} = 0$ as in four space-time dimensions and that 
the four-dimensional, non-zero trace,
\begin{equation}
{\rm Tr} [ \gamma^\mu \gamma^\nu \gamma^\rho \gamma^\sigma \gamma^5 ] = - 4 i \epsilon^{\mu\nu\rho\sigma},
\end{equation}
remains.  This FormCalc-based approach has been justified in~\cite{Chanowitz:1979zu}.  
In the second calculation, the treatment of 
$\gamma^5$ as described in~\cite{Larin:1993tq} is used, where
\begin{equation}
\gamma^5 = \frac{i}{4!} \epsilon_{\mu\nu\rho\sigma} \gamma^\mu \gamma^\nu \gamma^\rho \gamma^\sigma\; .
\end{equation}
Both approaches have been shown to give consistent results in our NLO calculation.

%% file: Section4_NumResults.tex

\section{Numerical results}
\label{sec_numres}

In this section we investigate the numerical influence of the SQCD one-loop corrections to both the 
neutral- and charged current \DY process with an associated hard jet.  
We examine these processes at the LHC with a \cm~energy of 7 TeV, unless noted otherwise.
We list the relevant input parameters in section~\ref{subsec_inputs}. 
The integrated cross section results and differential distributions for the \Zjet process are given in sections~\ref{subsec_Zy_crosssection} and \ref{subsec_Zy_distributions}, respectively.
The results for \Wjet production are discussed in sections~\ref{subsec_W_crosssection} and \ref{subsec_W_distributions}.

\subsection{Input parameters}
\label{subsec_inputs}

The SM input parameters are chosen in accordance with~\cite{Nakamura:2010zzi},
\begin{align}
\begin{split}
M_Z  = 91.1876~\text{GeV}, \qquad
M_W  &= 80.339~\text{GeV}, 
\\
\Gamma_Z  = 2.4952~\text{GeV}, \qquad~~
\Gamma_W  &= 2.085~\text{GeV}, 
\\
\cos\theta_W = M_W/M_Z,\qquad\quad~~
\alpha^{-1} &= \alpha(M_Z)^{-1} = 128.91,
\\
|V_{ud}| = 0.97428, \qquad |V_{us}| = 0.2253,
\qquad
|V_{cd}| &= 0.2252, \qquad |V_{cs}| = 0.97345.
\label{eq_inputs}
\end{split}
\end{align}
Quarks and leptons are considered massless and we do not further specify 
the flavor of the leptons in the final state as the SQCD corrections do not depend on the lepton flavor.
Factorization and renormalization scales are fixed and identified with the vector boson mass, 
$\mu_F = \mu_R = M_{Z,\,W}$. 
We use the central MSTW\,2008\,NLO ($68\%\,\text{CL}$) PDF set~\cite{Martin:2009iq}
in its LHAPDF implementation~\cite{Whalley:2005nh}, with the strong coupling $\alpha_s(\mu_R)$ 
they provide, yielding
\begin{align}
\alpha_s(M_Z) &= 0.12018, \qquad~
\alpha_s(M_W) = 0.12257.
\end{align} 

We require a set of basic kinematic cuts to be satisfied. 
As previously mentioned, a minimum \pT\ of the final state parton is required to render the cross section finite.  We also demand the two final-state leptons to have a minimum transverse momentum. 
Furthermore we require the leptons and the jet to be produced in the central region of the detector 
and apply a cut on their rapidities. 
For the \Zjet process we also require a minimum lepton-pair invariant mass, $\mll$.
We choose the following numerical values, 
\begin{align}
\begin{split}
\text{\Zjet}:\qquad
p_{T,\jet} &> 25\, \GeV, 
\qquad |y_{\jet}| < 2.5,
\\
p_{T,\ell^\pm} &> 25\, \GeV, 
\qquad |y_{\ell^\pm}| < 2.5,
\qquad \mll > 50\, \GeV\; . 
\\[1ex]
\text{\Wjet}:\qquad 
p_{T,\jet} &> 25\, \GeV, 
\qquad |y_{\jet}| < 2.5,
\\
p_{T,\ell}, ~\slash{\!\!\!p}_T &> 25\, \GeV, 
\qquad |y_{\ell}| < 2.5\; . 
\label{eq_cuts}
\end{split}
\end{align}
However, only the integrated cross sections depend strongly on the specific cuts chosen,   
while we have found that the relative importance of the SQCD corrections does not 
vary much when the cuts are tightened or loosened.

The only SUSY particles that appear in the loops are squarks and the gluino.
The SQCD corrections are flavor- and chirality-blind and   
no other SUSY parameters than the squark and gluino masses enter the calculation.
For simplicity we neglect the squark left-right mixing, set all squark masses equal and use 
a common squark mass, $\msquark$, and the gluino mass, $\mgluino$, as direct input.
There is no need to define a complete set of SUSY input parameters and 
we do not consider commonly used benchmark scenarios (as \eg\ SPS1a') here. 
As we will see below, this approach is sufficient for the purpose of our study
 to show the stability of \DYjet under the inclusion of SQCD corrections.
We use the following values for our numerical studies, if not otherwise noted,
\begin{align}
	\msquark = 600\,\GeV, \qquad 
	\mgluino = 500\,\GeV.
\end{align}
These sparticle masses are already at the lower limit of the mass region that is currently 
investigated by LHC experiments, see \eg~\cite{Collaboration:2011ida,Aad:2011xm},
and allow for a conservative estimate of the typical size of 
SQCD corrections to \DYjet processes.

\subsection{\Zjet integrated cross section results}
\label{subsec_Zy_crosssection}

Here we present the integrated cross section for charged dilepton production with a hard jet at the LHC.  
Table~\ref{table_Z_integratedCS_600-500} shows the
LO cross sections and SQCD contributions for the \qqbar\ and gluon-initiated partonic processes,
at a proton-proton \cm\ energy of $\sqrt{s}=7\,\TeV$ and $\sqrt{s}=14\,\TeV$.  

\TABULAR[t]{c|c|cc|c}{
\hline &&&\\[-2ex]
\Zjet & partonic  & LO & SQCD &  \\ [-1.5ex]
production & channel & cross section & contributions & \raisebox{2ex}{$\delta$}\\
\hline&&&\\[-2ex]
  $\sqrt{s} = 7$\,TeV 
&$q\bar{q}$ & 16.55 pb & 0.987 fb& 0.0059 \% \\[.5ex]
&$gq$ + $g\bar{q}$ & 37.73 pb & 2.520 fb& 0.0066 \% \\[.5ex]
&incl. & 54.27 pb & 3.507 fb& 0.0065 \% \\[.5ex]
\hline&&&\\[-2ex]
 $\sqrt{s} = 14$\,TeV
&$q\bar{q}$ & 31.70 pb & 1.947 fb& 0.0061  \% \\[.5ex]
&$gq$ + $g\bar{q}$ & 91.38 pb & 6.798 fb& 0.0074 \% \\[.5ex]
&incl. & 123.1 pb & 8.746 fb& 0.0071  \% \\[.5ex]
\hline
}
{Numerical results for integrated cross sections for the neutral-current \DY process
mediated by a $Z$ boson or virtual photon $\gamma^*$ 
in association with a hard jet at the LHC, with $\sqrt{s} = 7$\,TeV and  $\sqrt{s} = 14$\,TeV.
Shown are the leading order results in picobarn (pb), the SQCD contributions in femtobarn (fb) 
and the relative corrections $\delta$ for the partonic subchannels and the inclusive result (incl.). 
 Light quarks are implicitly summed over in the initial state, $q = u,d,c,s$.  
%
%
We consider $\msquark = 600\,$GeV, $\mgluino = 500\,$GeV and the cuts listed in \eqref{eq_cuts} have been applied.
Factorization and renormalization scale are set to $\mu=M_Z$ (with MSTW 2008 NLO) .   
\label{table_Z_integratedCS_600-500}
}

The prominent production modes at the LHC are the gluon-induced initial states, 
enhanced by the large gluon densities at small parton momentum fractions.
With the kinematic constraints in \eqref{eq_cuts}, the total 
integrated cross section is 54.27 pb at 7 TeV and 123.1 pb at 14 TeV ($q=u,d,c,s$).
The SQCD corrections, for squark and gluino masses of 600 and 500 GeV, respectively, 
account for an increase of $0.006\%-0.007\%$ in the inclusive integrated cross section 
as well as per partonic channel.   
The \cm\ energy at the LHC has little effect on the size of the corrections. 
Only the LO cross section strongly depends on the energy.
When going from 7 TeV to 14 TeV, 
the gluon-quark channels become even more important and 
we find that at LO they contribute 74\%.
However, the SQCD effects on the integrated cross section remain basically unchanged and  
small in absolute value.

\FIGURE[t]{
\includegraphics[width=.65\linewidth]{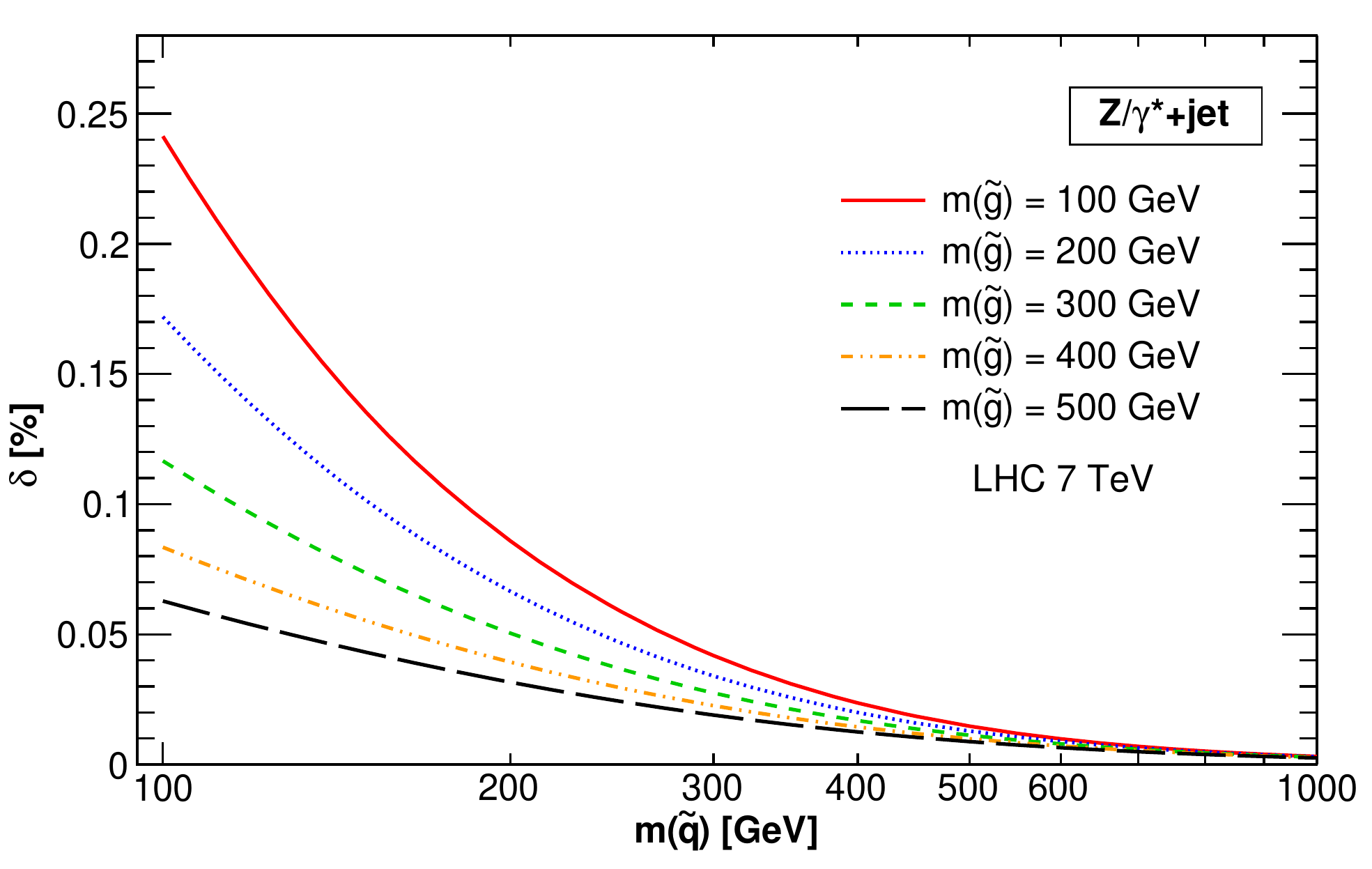}
\caption{Relative SQCD corrections of the integrated cross section, $\delta=(\sigma_{NLO}-\sigma_{LO})/\sigma_{LO}$, for charged
lepton-pair production mediated by a $Z$ boson or virtual photon in association with a hard jet at the LHC,
with $\sqrt{s} = 7\mbox{ TeV}$, as a function of a common squark mass,  $m_{\squark}$, 
for different gluino masses, $m_{\gluino}$. 
The cuts listed in \eqref{eq_cuts} have been applied.
\label{fig_susyscan}}
}

To further study the effects of SQCD corrections on the integrated cross section
of \Zjet production, we show in \figref{fig_susyscan} the relative corrections 
as a function of the common squark mass $\msquark$, for different values of the gluino mass.
As one would expect, 
the impact of the SQCD corrections increases as the SUSY particle masses decrease.  
However, even for very light squarks and gluinos of only 100 GeV   
the relative corrections are still significantly below the 1\% level.
\figref{fig_susyscan} shows that the \Zjet integrated cross section is stable under SQCD corrections, 
over a broad range of low-mass squarks and gluinos.   
For TeV-range SUSY particles the SQCD corrections are completely negligible.
This is comforting in the sense that if \Zjet is used as a normalization process, 
no SUSY BSM physics should be masked in this normalization.  
In terms of our calculation,
this stability also justifies our neglecting of squark left-right mixing, 
and use of a degenerate squark mass scheme.

\subsection{\Zjet kinematic distributions}
\label{subsec_Zy_distributions}

We have seen above that the SQCD contributions to the integrated cross section are subtle. 
However, the corrections can be more pronounced in
the differential distributions of kinematic variables. 
We consider the LO and NLO differential cross sections, and define the relative corrections $\delta$,
%
$\delta = \frac{\mathcal{O}_{NLO}-\mathcal{O}_{LO}}{\mathcal{O}_{LO}} \;$, 
for a given observable $\mathcal O$, where $\mathcal{O}_{NLO}$ is the sum of the LO contributions and the 
SQCD contributions.
We present distributions in the lepton-pair invariant mass, $\mll$,
the lepton transverse momentum and rapidity, $p_{T,\lep}$ and $y_{\lep}$, and
the jet transverse momentum and rapidity, $p_{T,\jet}$ and $y_{\jet}$.  
All distributions shown have been subject to the kinematic constraints given in
\eqref{eq_cuts}.

\renewcommand\plot[2]{\includegraphics[width=#1\linewidth]{plots/Zy_Aug24/#2}}

\FIGURE[t]{
\begin{minipage}{.49\linewidth}
\plot{1}{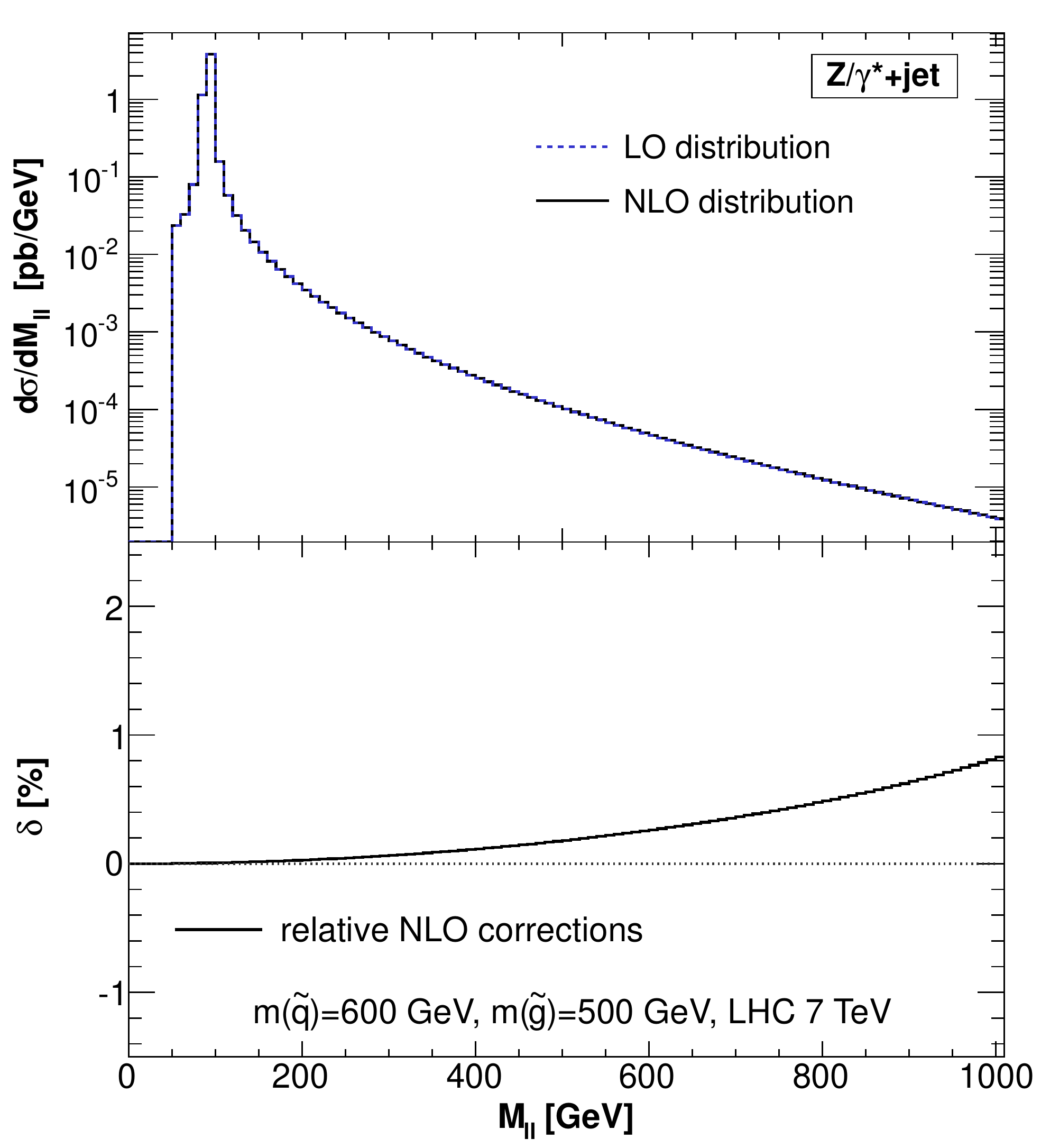}
\end{minipage}%
\begin{minipage}{.49\linewidth}
\plot{1}{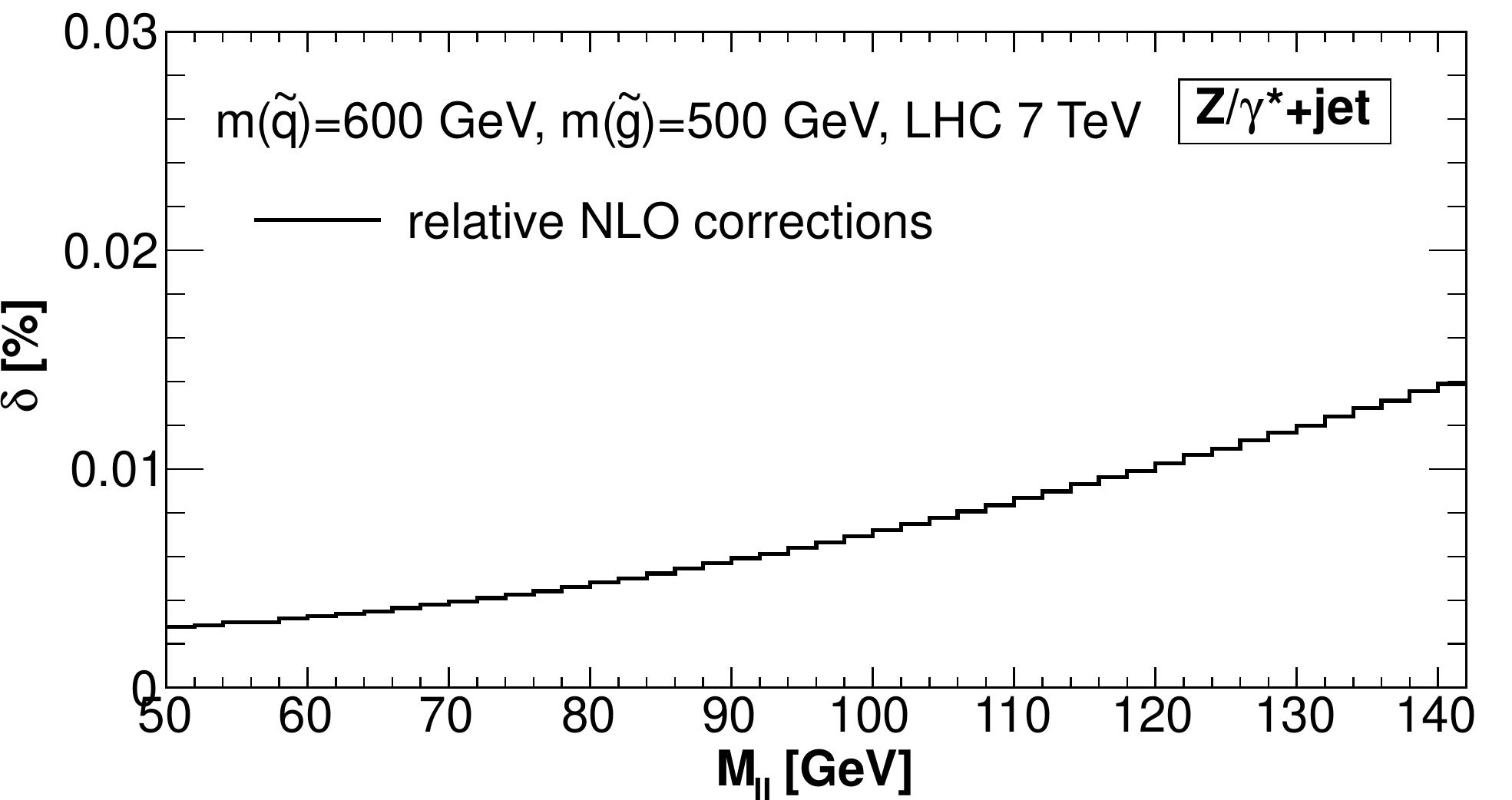}\\
\plot{1}{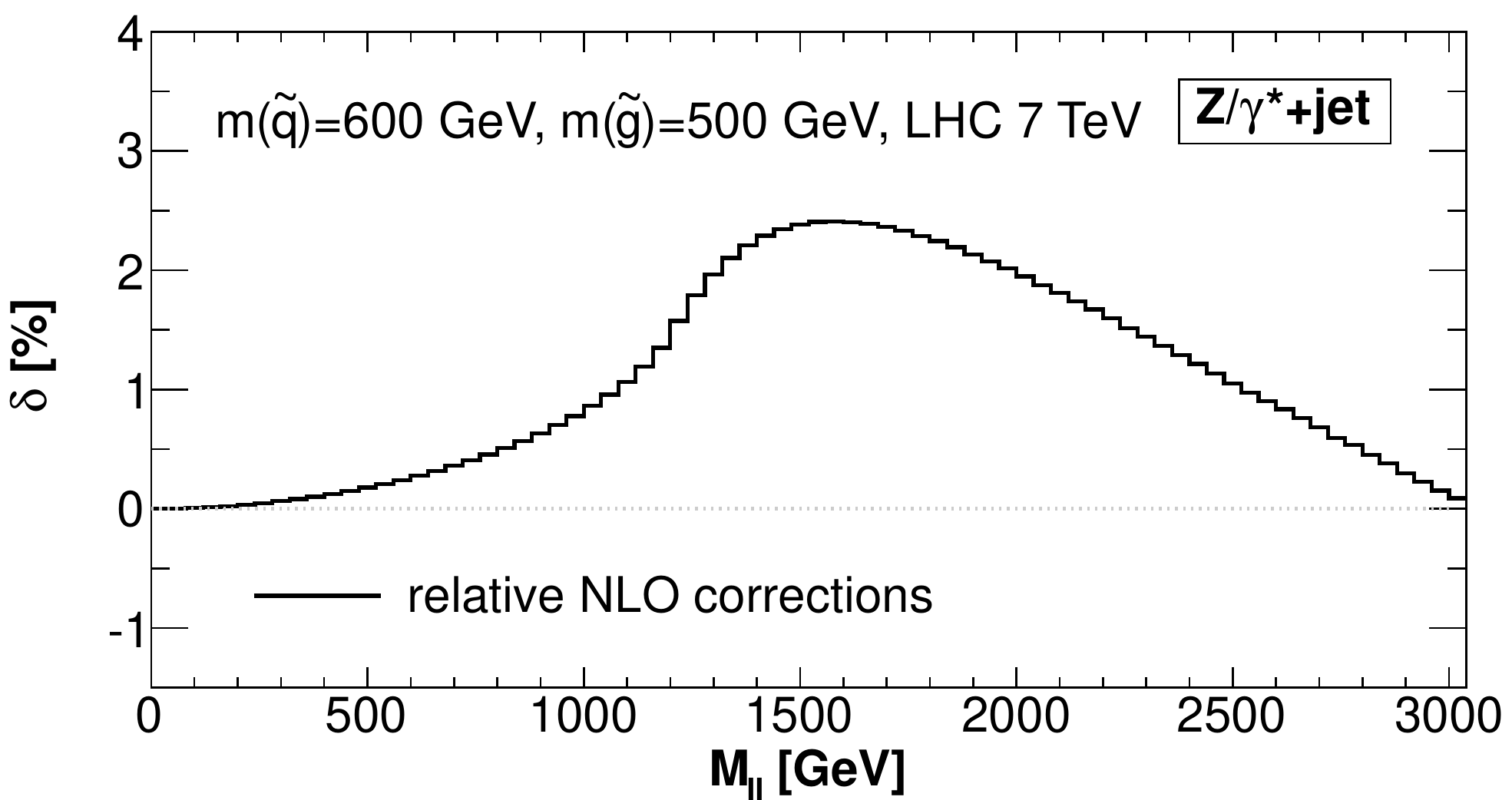}
\end{minipage}
\caption{Differential distributions for the neutral-current \DY process in assocation with a hard jet at the LHC.
NLO SQCD corrections are calculated for $\msquark = 600$\,GeV and $\mgluino = 500$\,GeV
and the cuts given in \eqref{eq_cuts} have been placed.
Shown are the absolute differential distributions for the LO and NLO processes
 (top left) and the relative difference between NLO and LO distributions (bottom left) 
with respect to lepton-pair invariant mass $\mll$. 
In the right panels, the relative corrections around the $Z$ boson resonance and in the high-$\mll$ region are shown, respectively.
\label{fig_Z_all_invMass}
}
}

In \figref{fig_Z_all_invMass}, 
distributions in the lepton-pair invariant mass $\mll$ are displayed.
We find that the SQCD corrections hardly affect the shape of the LO distribution and 
are below the $1\%$ level for $\mll < 1\,\TeV$ (left panels). 
In the vicinity of the $Z$~boson resonance, shown in the upper right panel, 
they are completely negligible and do not distort the SM result.  
For larger values of $\mll$ (lower right panel) 
the relative corrections can reach several percent, 
until a SUSY mass threshold is reached 
and then the relative corrections begin to fall.

\FIGURE[t]{
\plot{.49}{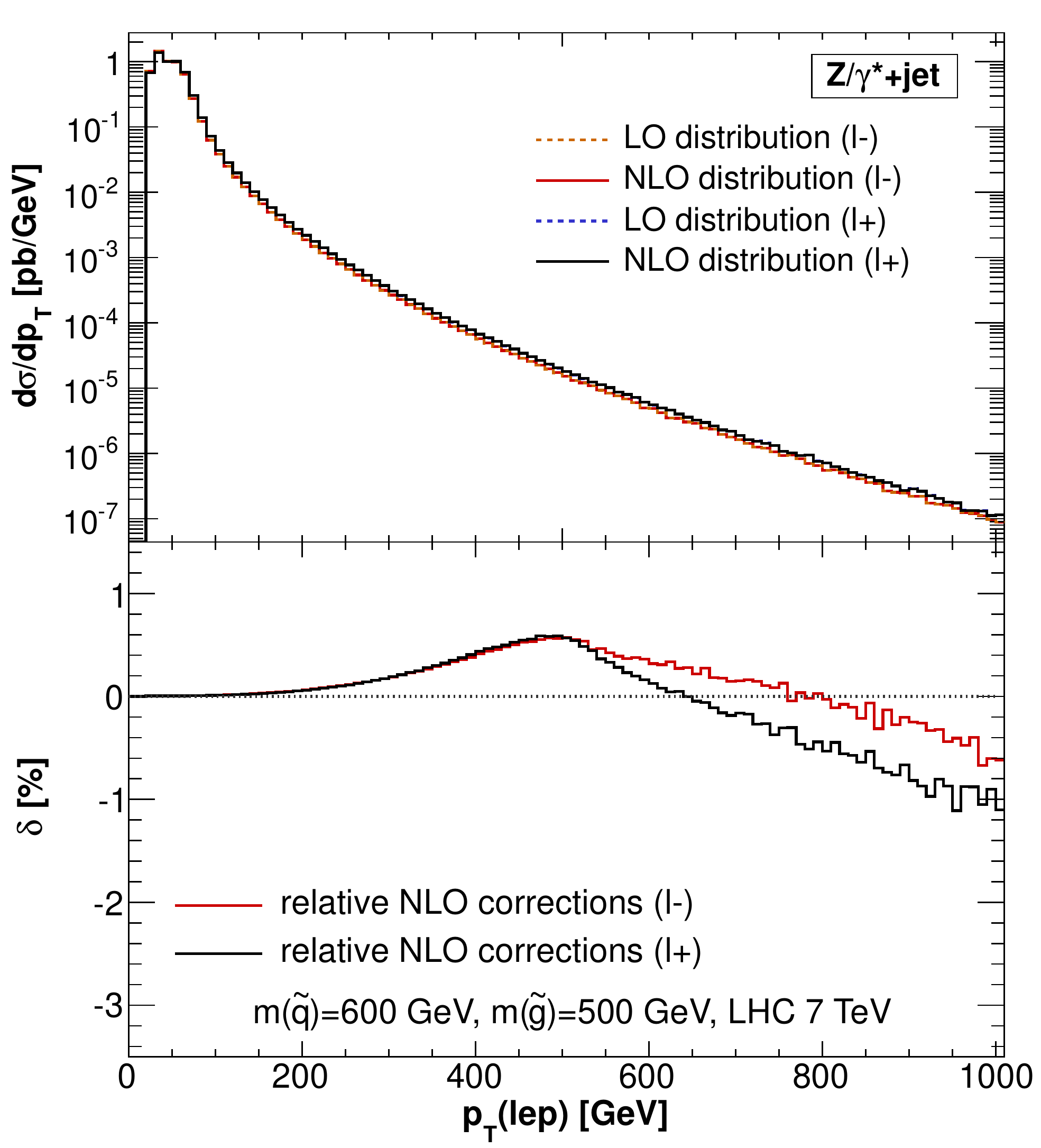}
\plot{.49}{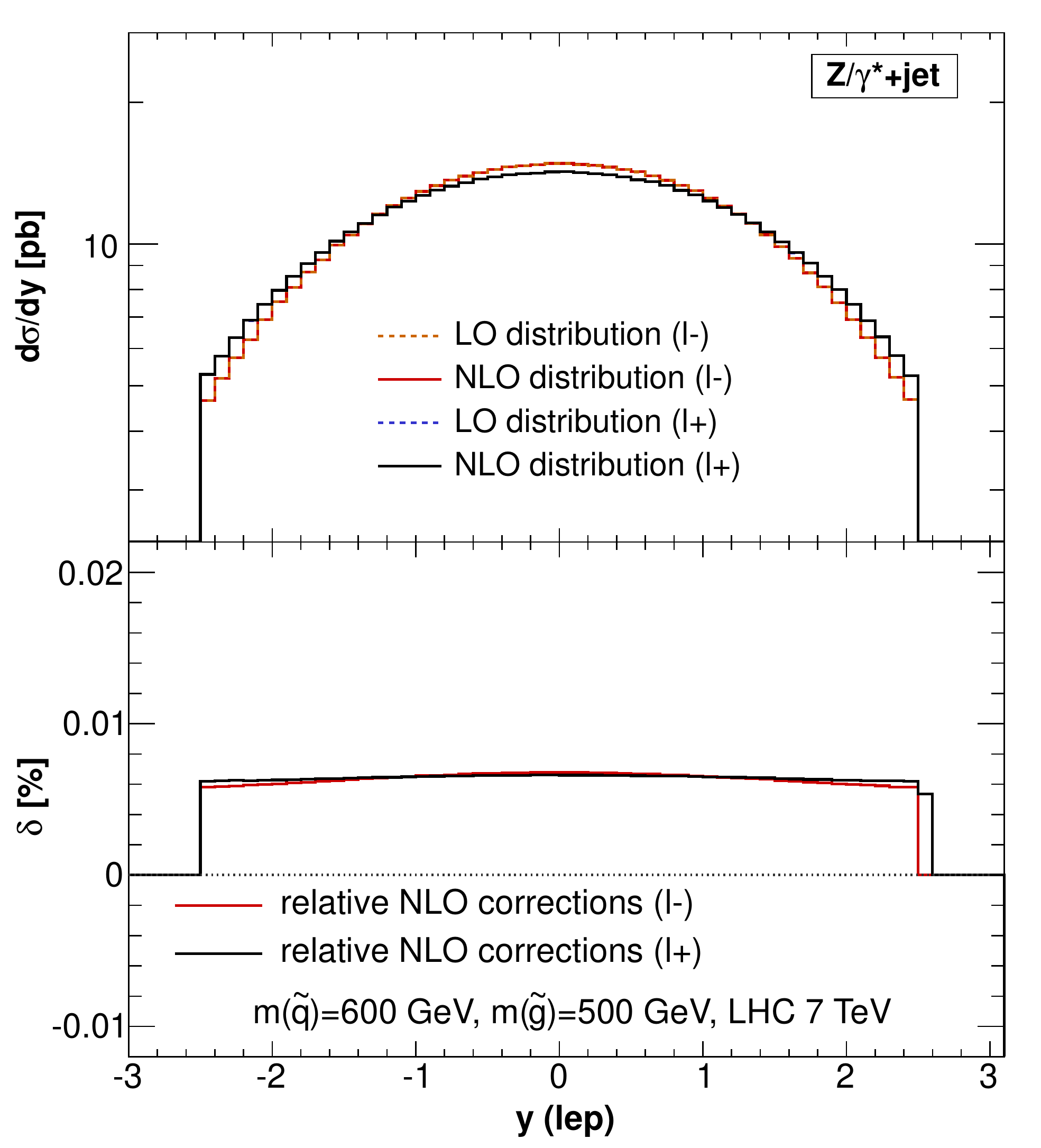}
\caption{Lepton transverse momentum, $p_{T,\lep}$, (a) and lepton rapidity, $y_{\lep}$, (b) 
distributions for the \Zjet process at the LHC with $\sqrt{s}=7\mbox{ TeV}$.  
The black line represents the lepton while the red line represents the antilepton. 
NLO SQCD corrections are calculated for $\msquark = 600$\,GeV and $\mgluino = 500$\,GeV
and the cuts given in \eqref{eq_cuts} have been placed.
\label{fig_Z_all_pTylep}
}
}

The lepton differential distributions in transverse momentum, $p_{T,\lep}$, and rapidity, $y_{\lep}$, 
are given in \figref{fig_Z_all_pTylep}.  
In the low- and intermediate-\pT\ range, the SQCD contributions to the $p_{T,\lep}$ distribution
are positive but small, increasing only to a maximum of about 1\% at around the threshold of the average sparticle mass.  In the high-\pT\ region, the relative corrections become non-negligible, on the order of a few negative percent.  
Concerning the lepton rapidity, \figref{fig_Z_all_pTylep}\,(right), we find that the relative corrections  
are nonzero but far below $1\%$ and thus do not have any impact on the shape of the SM-only distribution.
It is important to observe the discrepancies between the lepton and anti-lepton distributions in 
\figref{fig_Z_all_pTylep}. 
These differences occur already at the LO process and are a result of the $Z$ boson's axial coupling to fermions.  
Setting the axial coupling to zero would render the lepton and antilepton curves identical.

\FIGURE[t]{
\plot{.49}{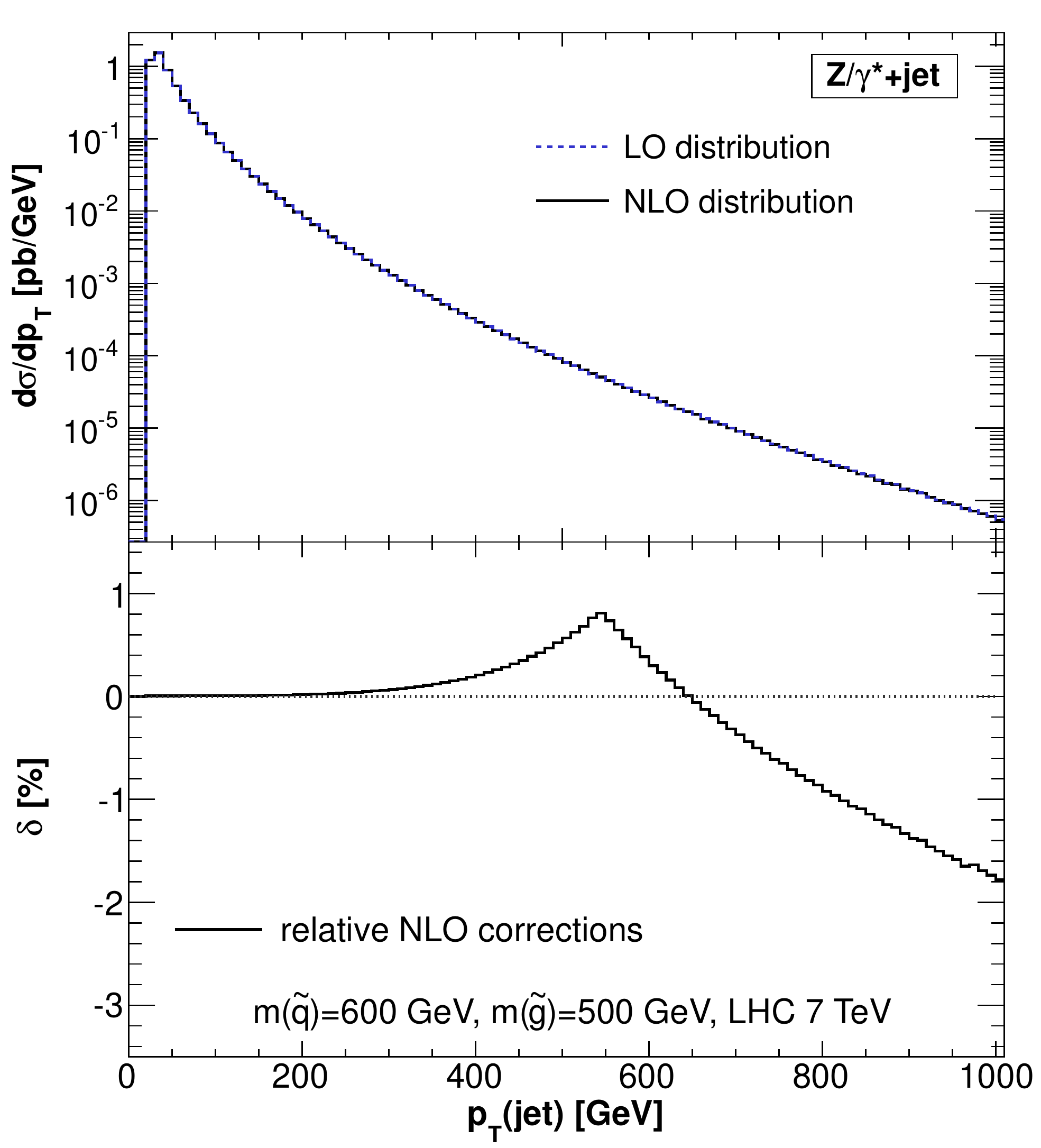}
\plot{.49}{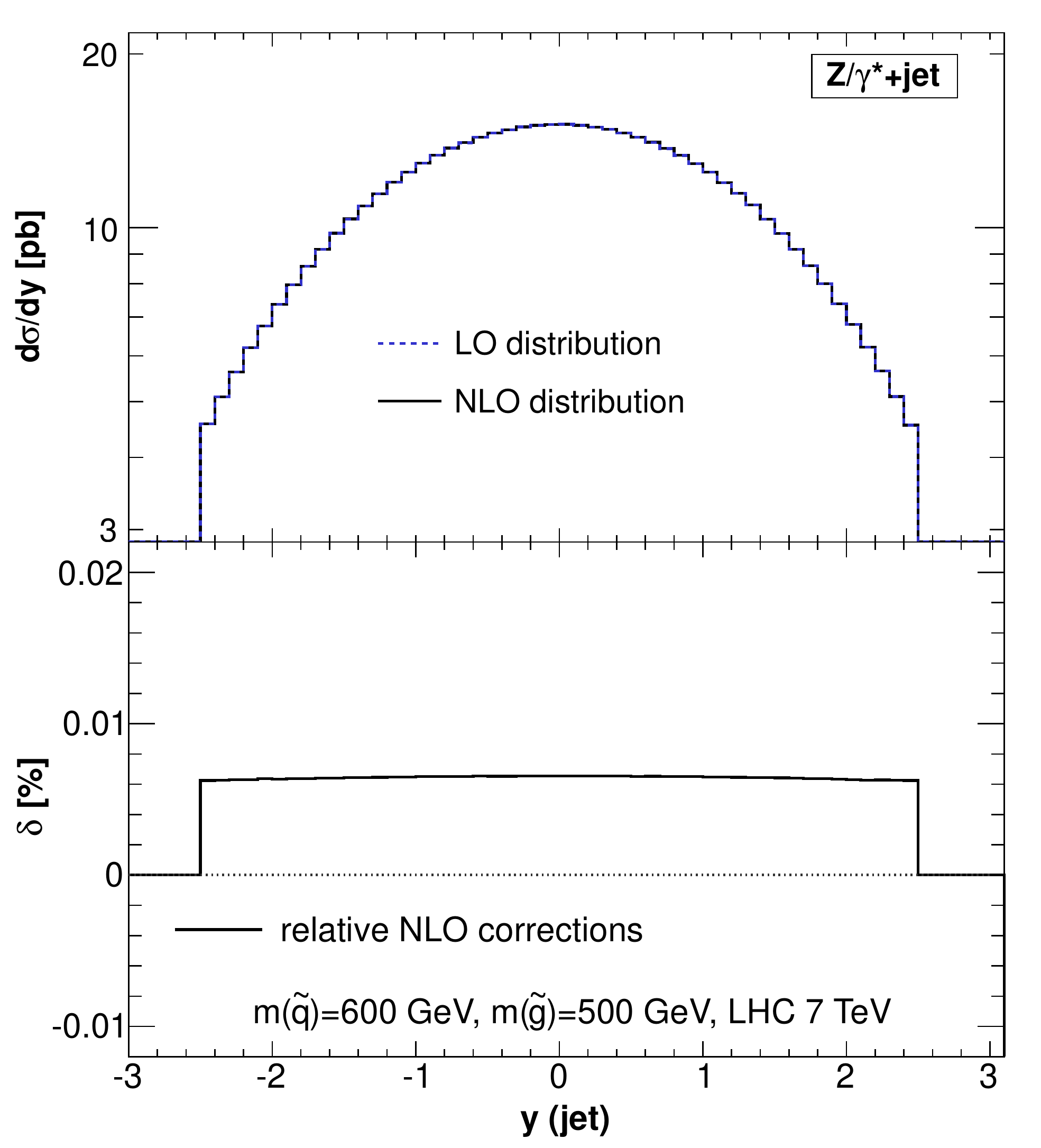}
\caption{Jet transverse momentum, $p_{T,\jet}$, (a) and jet rapidity, $y_{\jet}$, (b) 
distributions for the \Zjet process at the LHC with $\sqrt{s}=7\mbox{ TeV}$.
NLO SQCD corrections are calculated for $\msquark = 600$\,GeV and $\mgluino = 500$\,GeV
and the cuts given in \eqref{eq_cuts} have been placed.
\label{fig_Z_all_pTyjet}}
}

 The jet distributions, \figref{fig_Z_all_pTyjet}, exhibit the same behavior as the lepton distributions.
We see the threshold effect around $(\msquark + \mgluino)/2$ is even more pronounced 
and the relative corrections in the high-$p_{T,\jet}$ region (\pT\ $> 1.5\,\TeV$) contribute with roughly negative $4-5\%$, \figref{fig_Z_all_pTyjet}\,(left), 
while there are vanishing corrections to the jet rapidity, \figref{fig_Z_all_pTyjet}\,(right).

We also investigated differential distributions for lighter SUSY particles.
We found a very similar picture as described above, with the SUSY threshold shifted accordingly,
 and abstain from showing the plots here. 
For $\msquark = 400\,\GeV$ and $\mgluino = 350\,\GeV$,   
the SQCD corrections become only slightly more important, and reach the $1-2\%$ level at the sparticle thresholds and the $-5\%$ level for \pT\ $\approx 1500\,\GeV$.  

\afterpage{\clearpage}

%% file: Section4_NumResults_W.tex
\subsection{\Wjet integrated cross section results}
\label{subsec_W_crosssection}

In Table~\ref{table_W_integratedCS_600-500} we show
the results for the integrated cross section for lepton-neutrino production with a hard jet, 
mediated by a $W$ boson, at the LHC. 
The LO cross sections and SQCD contributions for the $q\bar{q}'$ and gluon-initiated partonic processes are given,
at a proton-proton \cm\ energy of $\sqrt{s}=7\,\TeV$ and $\sqrt{s}=14\,\TeV$.  
Similar to the \Zjet case, the SQCD corrections contribute between  
$0.004 - 0.006\%$ to the individual channels and the inclusive result. 
The gluon-induced processes are slightly more affected than the quark-annihilation channels.
The main cross section contribution comes from gluon-induced initial states, 
with increasing importance for higher \cm\ energies.
Thus the impact of the SQCD corrections on the inclusive cross section is somewhat enhanced  
at $14\,\TeV$ compared to $7\,\TeV$.

\TABULAR[t]{c|c|cc|c}{
\hline &&&\\[-2ex]
\Wjet & partonic  & LO & SQCD &  \\ [-1.5ex]
production & channel & cross section & contributions & \raisebox{2ex}{$\delta$}\\
\hline&&&\\[-2ex]
  $\sqrt{s} = 7$\,TeV 
&$q\bar{q}'$ & 129.7 pb& 5.746 fb& 0.0044 \% \\[.5ex]
&$gq$ + $g\bar{q}$ & 342.1 pb& 18.23 fb& 0.0053 \% \\[.5ex]
&incl. & 471.9 pb& 23.98 fb& 0.0051 \% \\[.5ex]
\hline&&&\\[-2ex]
 $\sqrt{s} = 14$\,TeV
&$q\bar{q}'$ & 245.3 pb& 10.81 fb& 0.0044 \% \\[.5ex]
&$gq$+ $g\bar{q}$ & 810.9 pb& 48.52 fb& 0.0060 \% \\[.5ex]
&incl. & 1056 pb& 59.33 fb& 0.0056 \% \\[.5ex]
\hline
}
{Numerical results for integrated cross sections for the charged-current \DY process
mediated by a $W$ boson in association with a hard jet at the LHC, with $\sqrt{s} = 7$\,TeV and  $\sqrt{s} = 14$\,TeV.
Shown are the leading order results in picobarn (pb), the SQCD contributions in femtobarn (fb) 
and the relative corrections $\delta$ for the partonic subchannels and the inclusive result (incl.). 
 Light quarks are implicitly summed over in the initial state, $q = u,d,c,s$.  
%
%
We consider $\msquark = 600\,$GeV, $\mgluino = 500\,$GeV and the cuts listed in \eqref{eq_cuts} have been applied.
Factorization and renormalization scale are set to $\mu=M_W$ (with MSTW 2008 NLO) .   
\label{table_W_integratedCS_600-500}
}

\subsection{\Wjet kinematic distributions}
\label{subsec_W_distributions}

The differential distributions for \Wjet production at the LHC with $\sqrt{s}=7\,\TeV$ are shown in 
Figs.~\ref{fig_W_all_invMass}, \ref{fig_W_all_pTylep}, and \ref{fig_W_all_pTyjet}.  They correspond to distributions in
transverse mass, lepton transverse momentum and rapidity, and jet transverse momentum and rapidity.
The transverse mass, $\MT$, and the \pT$_{\ell}$ distributions are particularly relevant for the measurement of the $W$ boson mass at hadron colliders. 
Here, the transverse mass is defined as 
$\MT = [( |p_{T,\ell}| + |\ptmiss| )^2 - ( \mathbf{p}_{T,\ell} + \mathbf{\slash \!\!\!p}_T )^2 ]^{1/2}$.

\renewcommand\plot[2]{\includegraphics[width=#1\linewidth]{plots/W_Sep01/#2}}


\FIGURE[t]{
\begin{minipage}{.49\linewidth}
\plot{1}{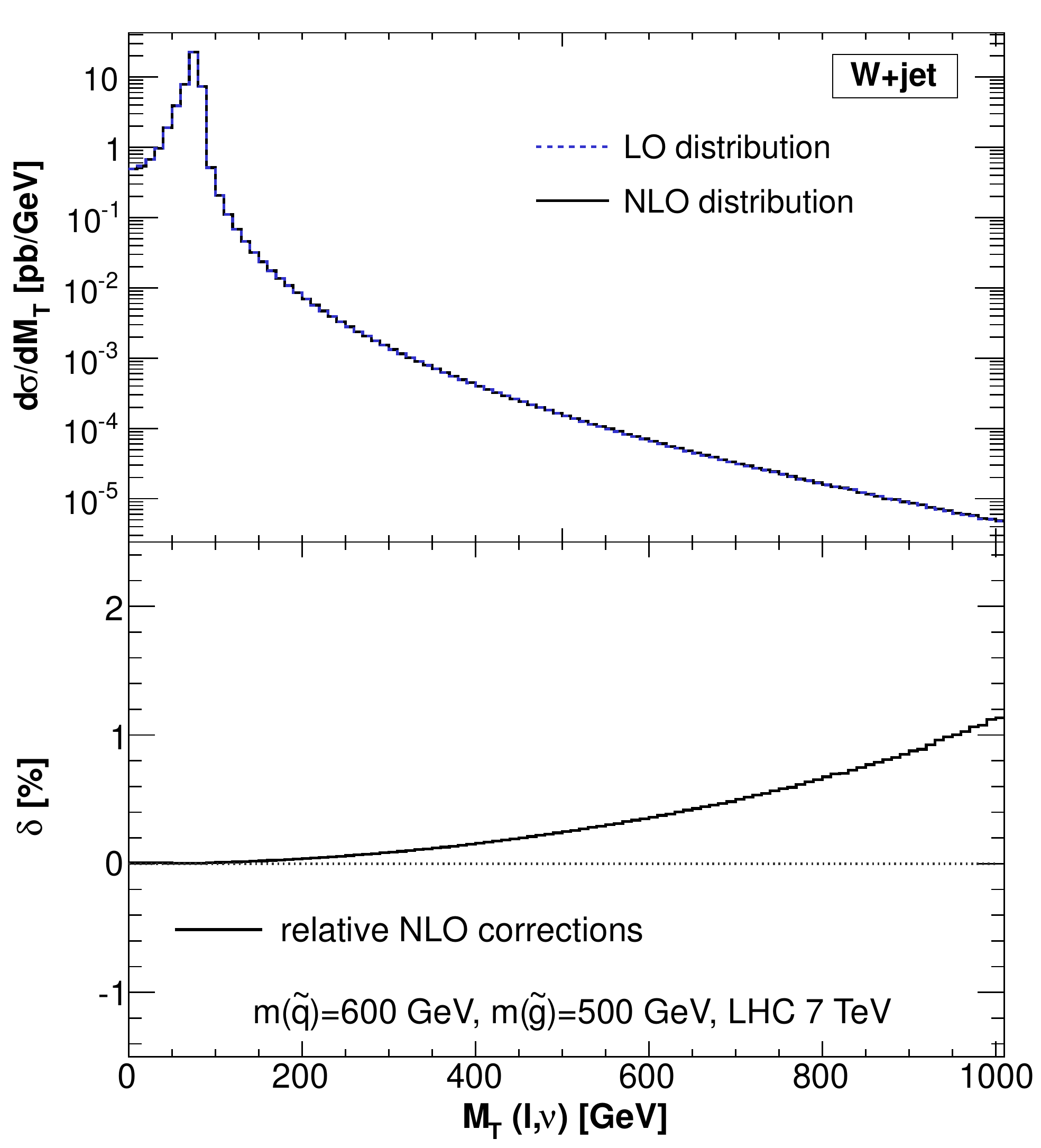}
\end{minipage}%
\begin{minipage}{.49\linewidth}
\plot{1}{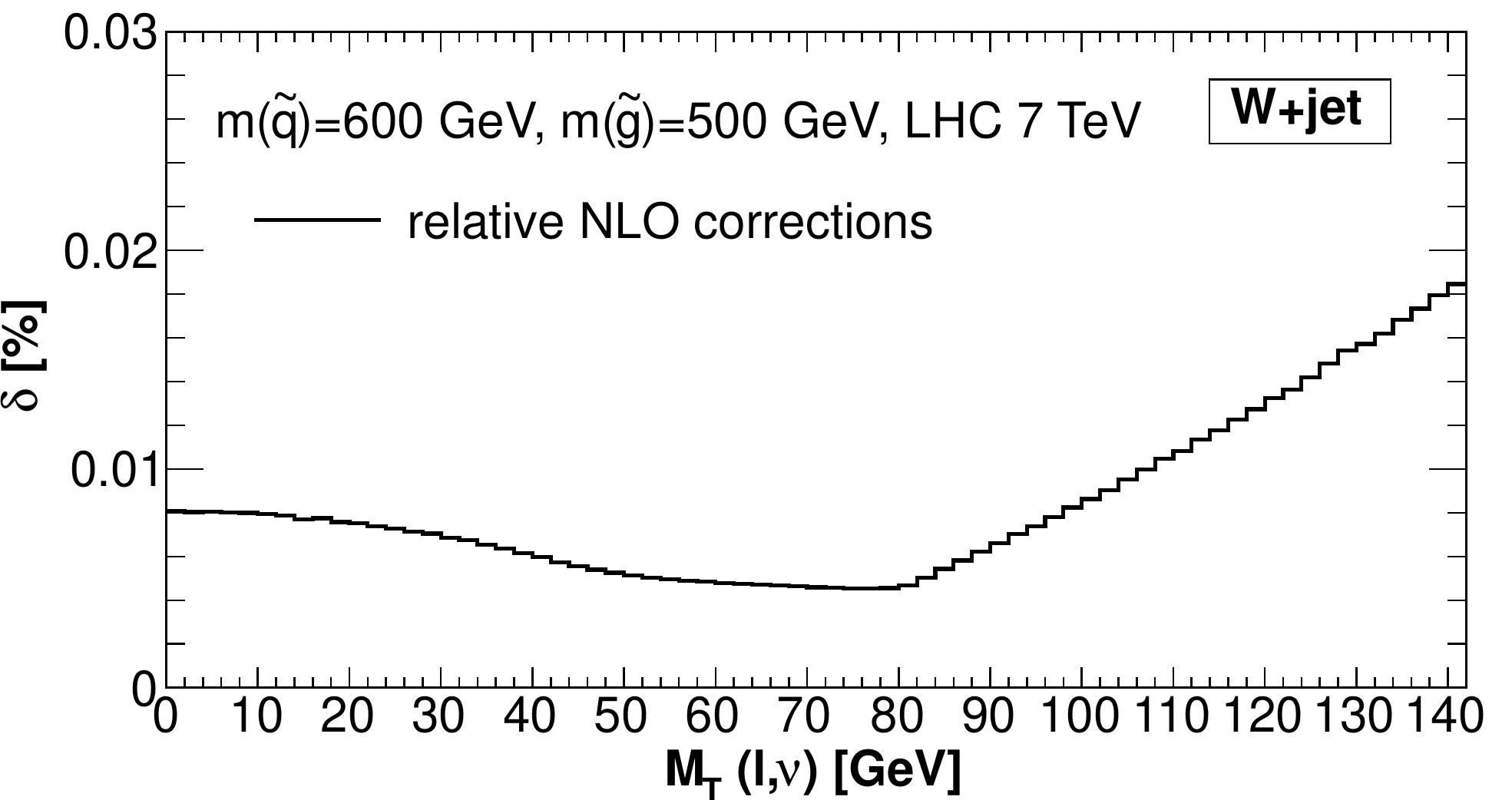}\\
\plot{1}{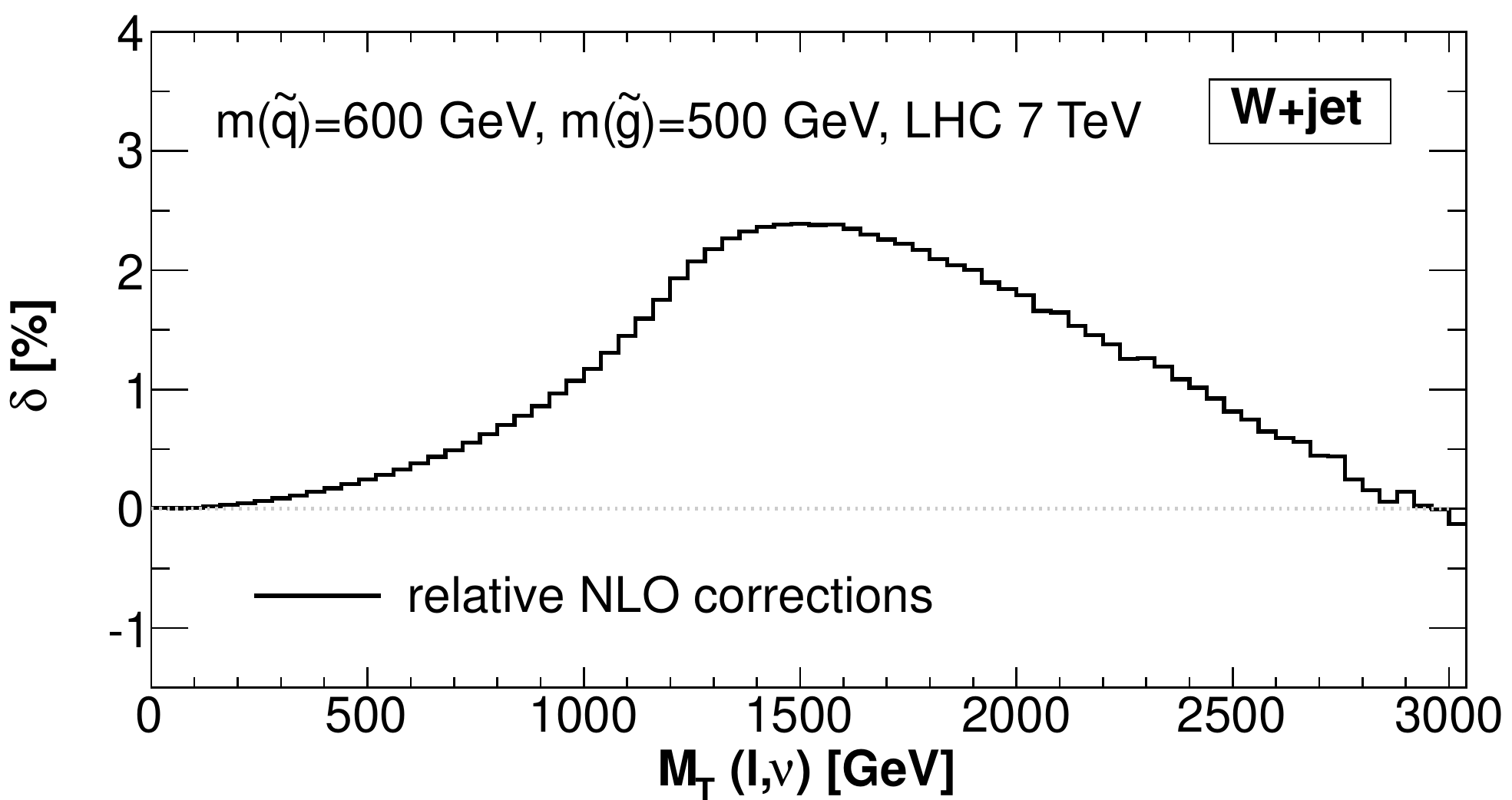}
\end{minipage}
\caption{Differential distributions for the charged-current \DY process in assocation with a hard jet at the LHC.
NLO SQCD corrections are calculated for $\msquark = 600$\,GeV and $\mgluino = 500$\,GeV
and the cuts given in \eqref{eq_cuts} have been placed.
Shown are the absolute differential distributions for the LO and NLO processes
 (top left) and the relative difference between NLO and LO distributions (bottom left) 
with respect to transverse mass $\MT$. 
In the right panels, the relative corrections around the $W$ boson resonance and in the high-$\MT$ region are shown, respectively.
\label{fig_W_all_invMass}
}
}

Again, the results are very similar to those for \Zjet production.
The $\MT$ distributions, displayed in \figref{fig_W_all_invMass}, receive 
percent-level corrections due to SQCD effects, which 
are minimal and almost vanishing when the $W$~boson is on-shell
and maximal around squark and gluino thresholds in the high-$\MT$ region.

\FIGURE[t]{
\plot{.49}{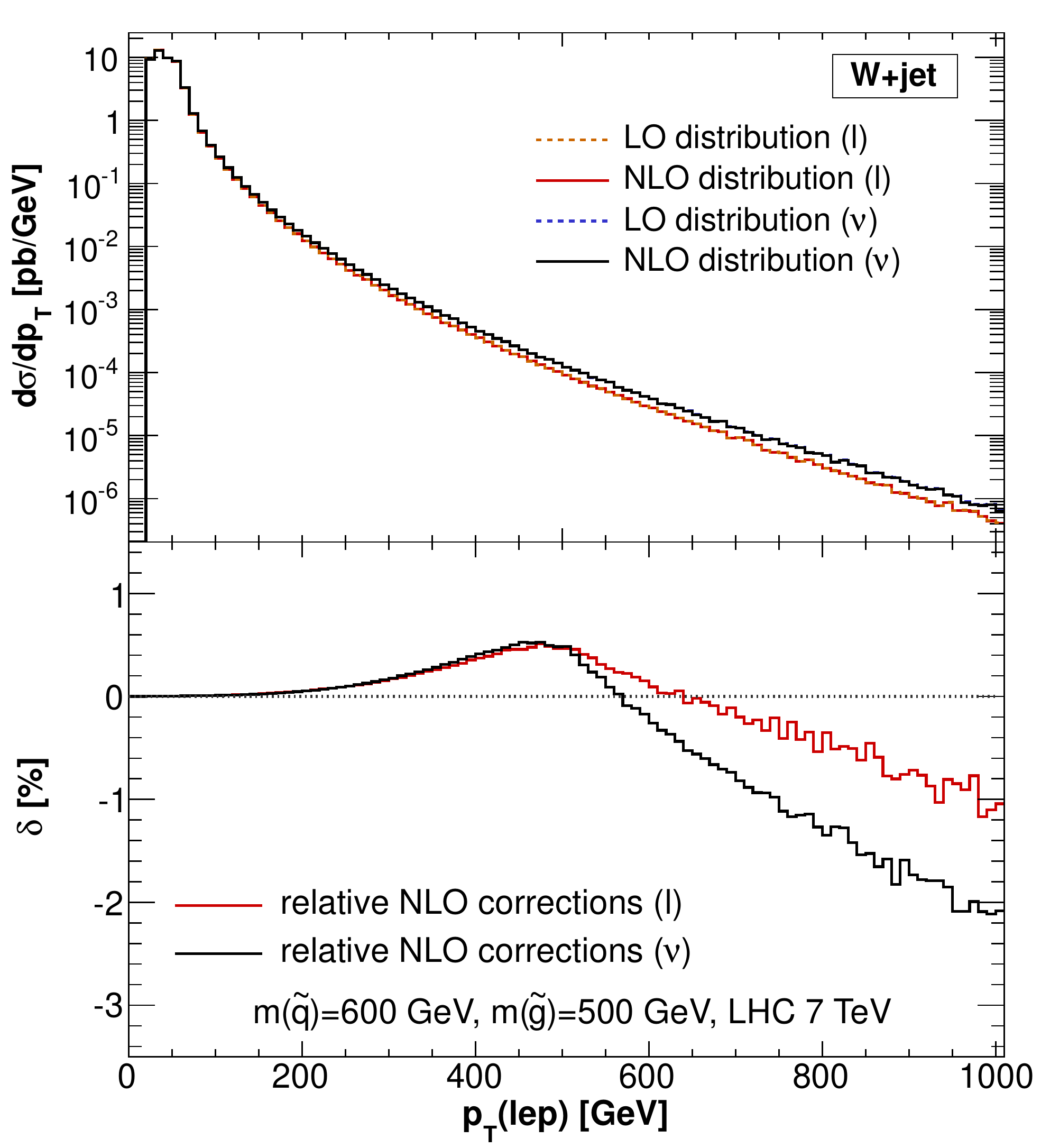}
\plot{.49}{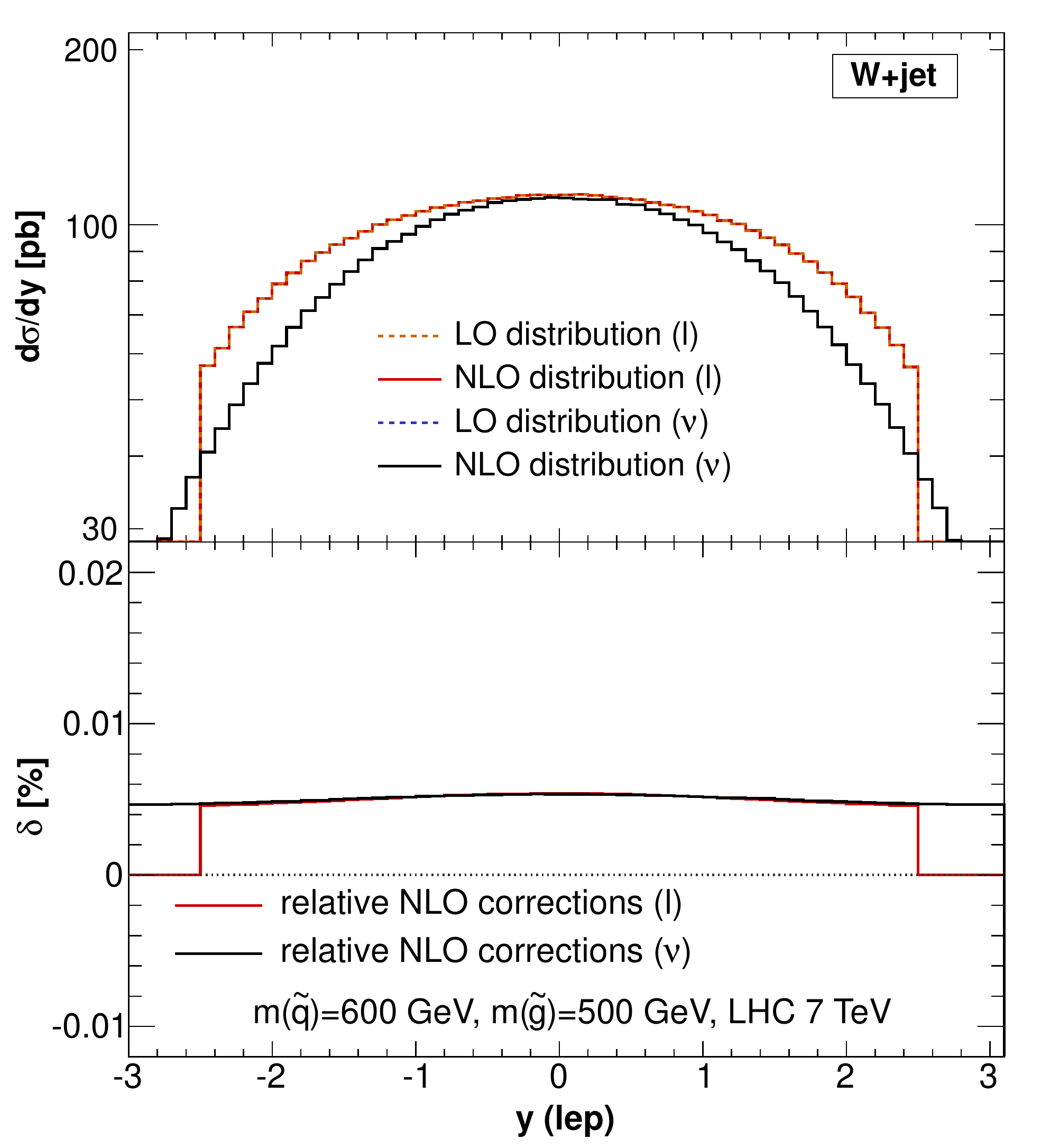}
\caption{Lepton transverse momentum, $p_{T,\lep}$, (a) and lepton rapidity, $y_{\lep}$, (b) distributions
for the \Wjet process at the LHC with $\sqrt{s}=7\mbox{ TeV}$.  
The red line represents the charged lepton 
while the black line represents the neutrino (missing transverse momentum, $\met$). 
\label{fig_W_all_pTylep}
}
}

The relative corrections to the $p_{T,\lep}$ and $p_{T,\jet}$ distributions, 
\figref{fig_W_all_pTylep} (left) and  \figref{fig_W_all_pTyjet} (left),
peak around the average sparticle mass where they amount to about $1\%$,
and grow negative in the high-\pT\ region, reaching a couple percent in the TeV region. 
The SQCD corrections in the lepton and jet rapidity distributions, 
\figref{fig_W_all_pTylep} (right) and  \figref{fig_W_all_pTyjet} (right),
are flat and can safely be neglected in an experimental analysis.

\clearpage

\FIGURE[t]{
\plot{.49}{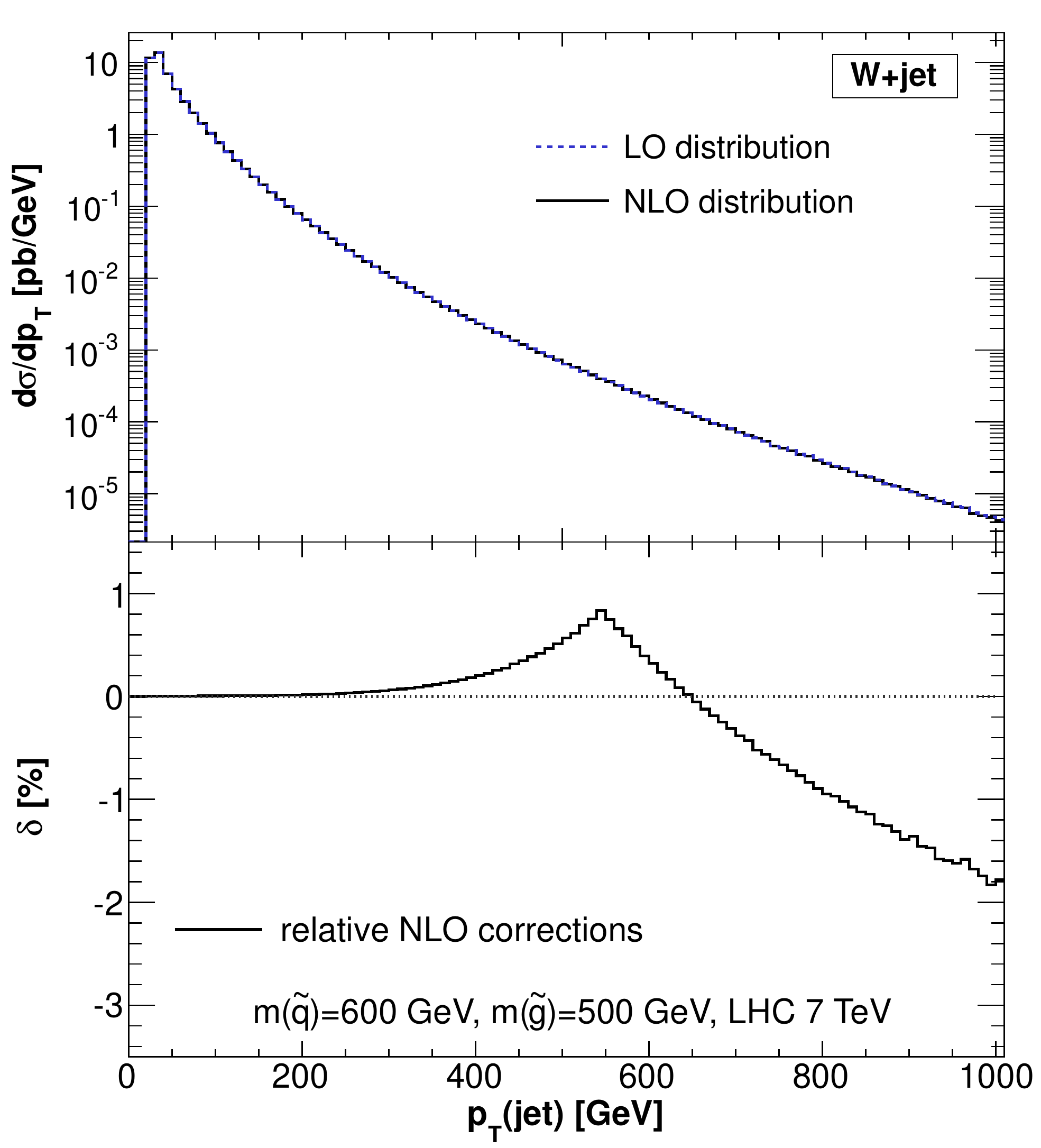}
\plot{.49}{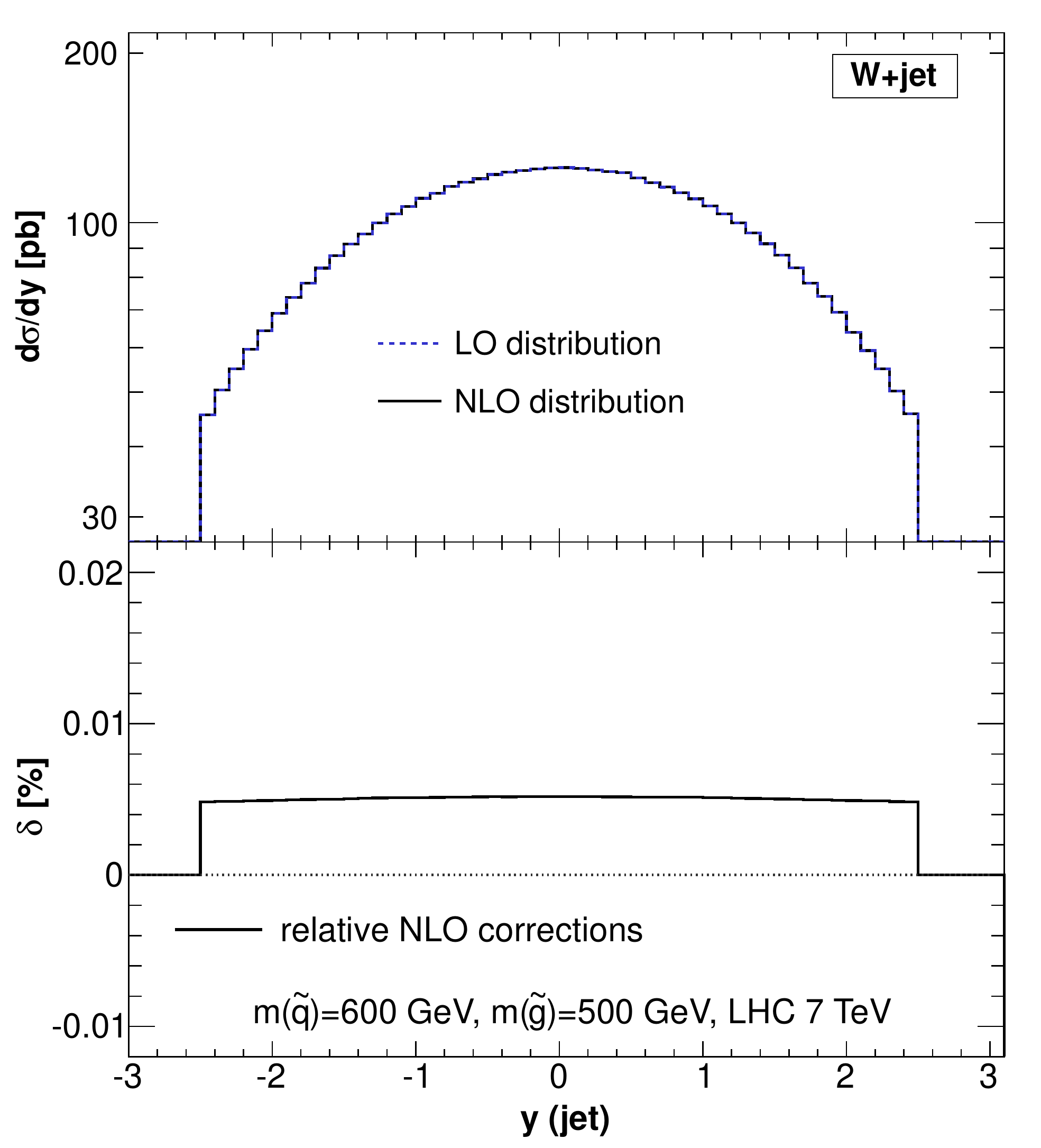}
\caption{Jet transverse momentum, $p_{T,\jet}$, (a) and jet rapidity, $y_{\jet}$, (b) distributions 
for the \Wjet process at the LHC with $\sqrt{s}=7\mbox{ TeV}$.  
\label{fig_W_all_pTyjet}}
}

%% file: Section5_Conclusions.tex
\section{Conclusions}
\label{sec_conclusion}

In this manuscript, we present a full treatment of the SQCD corrections to the neutral- and charged-current \DY process in 
association with a hard jet. We include the decay of the electroweak gauge bosons into a lepton pair or a lepton-neutrino pair, respectively, and take all off-shell effects due to the finite $Z$ and $W$ boson widths into account, as well as the contributions and interference effects of the photon-mediated diagrams in case of the neutral \DY process.   

Even though the \DY process, with an associated jet, can not be used in any practical 
manner to set indirect limits on SUSY particle masses or SUSY parameters, it is important to study the impact of 
SUSY (and other BSM physics) contributions to SM processes.  
As standard candles, and important backgrounds to new physics at the LHC, it is necessary to understand the stability of
EW gauge bosons production processes against BSM contributions.

We find that the relative corrections to 
the integrated cross section due to SQCD corrections are small, 
below the 0.5\% level even for very light SUSY particles.  
Examining the differential distributions in conventional
kinematic variables such as \pT, $\mll$, or $\MT$,
we find vanishing effects in the vicinity of the $Z$ or $W$ boson resonance peak,
while the relative SQCD corrections can increase to $2-5\%$  
when the electroweak gauge bosons are far off-shell.

%% file: Appendix_Renormalization.tex
\section{Counterterms and renormalization constants}
\label{app_renorm}

Here we list the counterterms for the renormalization of vertices and propagators in the SQCD one-loop 
amplitudes for \DYjet production.  
All quarks are considered massless.

By using multiplicative renormalization, we replace in the QCD Lagrangian the 
left- and right-handed quark fields, $\Psi_{L,R}$, 
the gluon field, $G^a$,
and the strong coupling constant, $g_s$, with
\begin{equation}
\Psi_{L,R} \rightarrow \sqrt{Z_{L,R}}\, \Psi_{L,R} \;,\qquad 
G^a_\mu \rightarrow \sqrt{Z_G}\, G^a_\mu \; , \qquad 
g_s \rightarrow Z_{g_s} \, g_s \; .
\end{equation}
Expanding $Z_i = 1 + \delta Z_i$, 
and introducing $\delta Z_{V,A} = \frac{1}{2}(\delta Z_L \pm \delta Z_R)$,
we find the following Feynman rules for the self-energy and vertex counterterms that 
are relevant in our calculation.
\begin{eqnarray}
\begin{array}{c}
\includegraphics{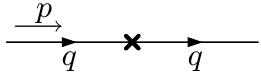}
\end{array}
&:&\ i  \!\! \not\! p \, (\delta Z_V-\delta Z_A \gamma_5), \\[2ex]
\begin{array}{c}
\includegraphics{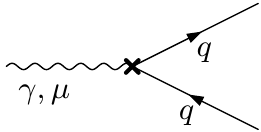}
\end{array}
&:& -i e e_q \, \gamma^\mu  \, (\delta Z_V-\delta Z_A \gamma_5), \\[2ex] 
\begin{array}{c}
\includegraphics{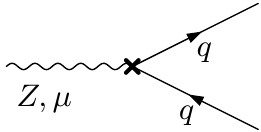}
\end{array}
&:& -\frac{i e}{c_W s_W}  \, \gamma^\mu \, (g_V^q - g_A^q \gamma_5)\, (\delta Z_V-\delta Z_A \gamma_5),
\qquad 
\begin{array}{l}
g_V^q = \frac{1}{2}I_3^q - e_q s_W^2, \\
g_A^q = \frac{1}{2} I_3^q,
\end{array}\qquad\quad
 \\[2ex] 
\begin{array}{c}
\includegraphics{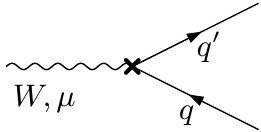}
\end{array}
&:& -\frac{i e}{2 \sqrt{2} s_W} \, V^{\text{CKM}}_{qq'} \, \gamma^\mu \, (1-\gamma_5)\, (\delta Z_V+\delta Z_A), \\[2ex] 
\begin{array}{c}
\includegraphics{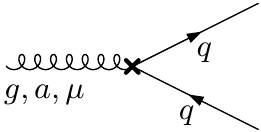}
\end{array}
&:& -i g_s T^a \, \gamma^\mu \, (\delta Z_V-\delta Z_A \gamma_5 + \delta Z_{g_s} + \frac{1}{2} \delta Z_G ),
\label{eq_CT_gqq}
\end{eqnarray}
where $e_q$ is the electric charge of quark $q$ and  
$I_3^q$ denotes the eigenvalue of the third component of its weak isospin.
$T^a$, $a=1..8$, are the color matrices of $SU(3)_C$, while we have omitted the color indices of the quarks.
We use the abbreviation $s_W$ and $c_W$ for the sine and cosine of the electroweak mixing angle $\theta_W$. 
As usual, $q$ and $q'$ are quarks of opposite isospin and  
$V^{\text{CKM}}_{qq'}$ is the corresponding entry of the CKM quark mixing matrix.
 
For the renormalization of the gluon field and the strong coupling we use the
$\overline{MS}$ scheme, modified to decouple the squarks and the gluino \cite{Collins:1978wz,Nason:1989zy}.  
The renormalization constant of the strong coupling is then given by, see also \cite{Berge:2007dz},
\begin{align}
   \delta Z_{g_s} &= - \frac{\alpha_s}{4\pi} \left\lbrace
	\frac{\beta_0}{2} \Delta + 
	\frac{1}{12} \sum_{\squark} \log\left( \frac{\msquark^2}{\mu^2} \right) +
	\log \left( \frac{\mgluino^2}{\mu^2} \right)
	\right\rbrace,
\end{align}
where the sum runs over the twelve squark eigenstates, $\mu$ is the renormalization 
scale, $\Delta = 1/\epsilon - \gamma_E + \ln4\pi$; and  
\begin{align}
\beta_0 &=  -\frac{2}{3} N  -\frac{1}{3} n_{\squark},
\end{align}
with $N=3$ and $n_{\squark} = 6$ for the six squark flavors, includes only the gluino and 
squark contributions.   
In this case, 
there is a simple relation between the gluon field and strong coupling renormalization constants, 
\begin{align}
	\delta Z_G = -2 \, \delta Z_{g_s},
\end{align}
and they actually cancel out in \eqref{eq_CT_gqq} and do not enter our calculation.

In the quark sector, we use the on-shell scheme to fix the renormalization constants $\delta Z_{V,A}$ of the (massless) quarks, 
\ie\ the renormalization constants are obtained by evaluating the vector and axial components of the 
quark self-energy at the on-shell quark mass,
\begin{align}
\delta Z_{V,A}  &= -\Sigma_{V,A}(p^2=0).
\end{align}
The SQCD corrections to the quark self-energy consist of a squark-gluino bubble insertion in the quark line
and can be expressed in the following general, compact form~\cite{Berge:2007dz},
\begin{align}
\begin{split}
\Sigma_V(p^2) &= - \frac{\alpha_s}{4\pi} \sum_i \left( \left(g_S^i\right)^2 + \left(g_P^i\right)^2 \right)\,
	\frac{2}{3} B_1(p^2,m_{\tilde{g}},m_{\tilde{q}_i})\; ,\\
\Sigma_A(p^2) &= \frac{\alpha_s}{4\pi} \sum_i 2 g_S^i g_P^i \,
	\frac{2}{3} B_1(p^2,m_{\tilde{g}},m_{\tilde{q}_i})\; .
\end{split}
\end{align}
Here, $B_1(p^2,m_1^2,m_2^2)$ is the two-point function as defined in~\cite{Hahn:1998yk}.
$i$ sums over the two squark mass eigenstates and 
$g_S^i$ and $g_P^i$ are scalar and pseudoscalar couplings.
In the limit of no left-right mixing in the squark sector, $g_S^{L,R} = \pm 1$ and $g_P^{L,R} = 1$.
When considering the mixing, then
\begin{align}
\begin{split}
	g_S^{1,2} &= \cos\theta_{\squark} \, g_S^{L,R} \pm 
		\sin\theta_{\squark} \, g_S^{R,L},
\\
	g_P^{1,2} &= \cos\theta_{\squark} \, g_P^{L,R} \pm 
		\sin\theta_{\squark} \, g_P^{R,L},
\end{split}
\end{align}
where $\theta_{\squark}$ is the squark mixing angle (notation as in \cite{Berge:2007dz}).

In our numerical study, we neglect the left-right squark mixing, and consider degenerate squark masses, 
$m_{\tilde{q}_1} = m_{\tilde{q}_2} = \msquark$.
In this limit, we find $\Sigma_A(p^2) = 0$ and the only renormalization constant 
that remains is $\delta Z_V$, with
\begin{align}
\begin{split}
\delta Z_V =&  - \frac{\alpha_s}{3 \pi} 
 \left\lbrace 
	 \Delta +  
	\frac{3 \mgluino^2 - \msquark^2 }{2(m_{\tilde{g}}^2-m_{\tilde{q}}^2)} -
 	\frac{\mgluino^4}{(m_{\tilde{g}}^2-m_{\tilde{q}}^2)^2}
	 	\ln\left( \frac{m_{\tilde{g}}^2}{\mu_R^2} \right) + 
	\frac{\msquark^2 ( 2 \mgluino^2-\msquark^2)}{(m_{\tilde{g}}^2-m_{\tilde{q}}^2)^2}
		\ln \! \left(\frac{m_{\tilde{q}}^2}{\mu_R^2}\right)  
  \right\rbrace.
\end{split}
\end{align}

%% file: main.bbl
\providecommand{\href}[2]{#2}\begingroup\raggedright\begin{thebibliography}{10}

\bibitem{Collaboration:2011cm}
CMS, {\it {Measurement of the Drell-Yan Cross Section in pp Collisions at
  sqrt(s) = 7 TeV}},  \href{http://xxx.lanl.gov/abs/1108.0566}{{\tt
  arXiv:1108.0566}}.

\bibitem{Collaboration:2011nx}
{\bf CMS} Collaboration, {\it {Measurement of the Inclusive W and Z Production
  Cross Sections in pp Collisions at sqrt(s) = 7 TeV}},
  \href{http://xxx.lanl.gov/abs/1107.4789}{{\tt arXiv:1107.4789}}.

\bibitem{Collaboration:2011gj}
{\bf ATLAS} Collaboration, {\it {Measurement of the transverse momentum
  distribution of Z/gamma* bosons in proton-proton collisions at sqrt(s)=7 TeV
  with the ATLAS detector}},  \href{http://xxx.lanl.gov/abs/1107.2381}{{\tt
  arXiv:1107.2381}}.

\bibitem{Aad:2010yt}
{\bf ATLAS} Collaboration, G.~Aad {\em et.~al.}, {\it {Measurement of the W $\to$
  lnu and Z/gamma* $\to$ ll production cross sections in proton-proton collisions
  at sqrt(s) = 7 TeV with the ATLAS detector}},  {\em JHEP} {\bf 1012} (2010)
  060, [\href{http://xxx.lanl.gov/abs/1010.2130}{{\tt arXiv:1010.2130}}].

\bibitem{Ball:2010de}
R.~D. Ball, L.~Del~Debbio, S.~Forte, A.~Guffanti, J.~I. Latorre, {\em et.~al.},
  {\it {A first unbiased global NLO determination of parton distributions and
  their uncertainties}},  {\em Nucl.Phys.} {\bf B838} (2010) 136--206,
  [\href{http://xxx.lanl.gov/abs/1002.4407}{{\tt arXiv:1002.4407}}].

\bibitem{Thorne:2010kj}
R.~Thorne, A.~Martin, W.~Stirling, and G.~Watt, {\it {The Effects of combined
  HERA and recent Tevatron W $\to$ l $\nu$ charge asymmetry data on the MSTW
  PDFs}},  {\em PoS} {\bf DIS2010} (2010) 052,
  [\href{http://xxx.lanl.gov/abs/1006.2753}{{\tt arXiv:1006.2753}}].

\bibitem{Lai:2010vv}
H.-L. Lai, M.~Guzzi, J.~Huston, Z.~Li, P.~M. Nadolsky, {\em et.~al.}, {\it {New
  parton distributions for collider physics}},  {\em Phys.Rev.} {\bf D82}
  (2010) 074024, [\href{http://xxx.lanl.gov/abs/1007.2241}{{\tt
  arXiv:1007.2241}}].

\bibitem{Dittmar:1997md}
M.~Dittmar, F.~Pauss, and D.~Zurcher, {\it {Towards a precise parton luminosity
  determination at the CERN LHC}},  {\em Phys.Rev.} {\bf D56} (1997)
  7284--7290, [\href{http://xxx.lanl.gov/abs/hep-ex/9705004}{{\tt
  hep-ex/9705004}}].

\bibitem{Rizzo:2006nw}
T.~G. Rizzo, {\it {$Z^\prime$ phenomenology and the LHC}},
  \href{http://xxx.lanl.gov/abs/hep-ph/0610104}{{\tt hep-ph/0610104}}.

\bibitem{Chiang:2011kq}
C.-W. Chiang, N.~D. Christensen, G.-J. Ding, and T.~Han, {\it {Discovery in
  Drell-Yan Processes at the LHC}},
  \href{http://xxx.lanl.gov/abs/1107.5830}{{\tt arXiv:1107.5830}}.

\bibitem{ATLAS-TDR:1999fr}
{\bf ATLAS} Collaboration, {\it {Detector and physics performance technical
  design report. Volume 2}},  \href{http://xxx.lanl.gov/abs/ATLAS-TDR-15}{{\tt
  ATLAS-TDR-15}}.

\bibitem{Ball:2007zza}
{\bf CMS} Collaboration, G.~Bayatian {\em et.~al.}, {\it {CMS technical design
  report, volume II: Physics performance}},  {\em J.Phys.G} {\bf G34} (2007)
  995--1579.

\bibitem{systematics}
G.~Dissertori, {\it {Prospects for measuring hard processes the LHC}}, . Talk
  presented at the ${\rm HP}^2$ Workshop, Zurich, Switzerland (2006).

\bibitem{Hamberg:1990np}
R.~Hamberg, W.~van Neerven, and T.~Matsuura, {\it {A Complete calculation of
  the order $\alpha_s^{2}$ correction to the Drell-Yan $K$ factor}},  {\em
  Nucl.Phys.} {\bf B359} (1991) 343--405.

\bibitem{Harlander:2002wh}
R.~V. Harlander and W.~B. Kilgore, {\it {Next-to-next-to-leading order Higgs
  production at hadron colliders}},  {\em Phys.Rev.Lett.} {\bf 88} (2002)
  201801, [\href{http://xxx.lanl.gov/abs/hep-ph/0201206}{{\tt
  hep-ph/0201206}}].

\bibitem{Anastasiou:2003yy}
C.~Anastasiou, L.~J. Dixon, K.~Melnikov, and F.~Petriello, {\it {Dilepton
  rapidity distribution in the Drell-Yan process at NNLO in QCD}},  {\em
  Phys.Rev.Lett.} {\bf 91} (2003) 182002,
  [\href{http://xxx.lanl.gov/abs/hep-ph/0306192}{{\tt hep-ph/0306192}}].

\bibitem{Anastasiou:2003ds}
C.~Anastasiou, L.~J. Dixon, K.~Melnikov, and F.~Petriello, {\it {High precision
  QCD at hadron colliders: Electroweak gauge boson rapidity distributions at
  NNLO}},  {\em Phys.Rev.} {\bf D69} (2004) 094008,
  [\href{http://xxx.lanl.gov/abs/hep-ph/0312266}{{\tt hep-ph/0312266}}].

\bibitem{Melnikov:2006di}
K.~Melnikov and F.~Petriello, {\it {The $W$ boson production cross section at
  the LHC through $O(\alpha^2_s)$}},  {\em Phys.Rev.Lett.} {\bf 96} (2006)
  231803, [\href{http://xxx.lanl.gov/abs/hep-ph/0603182}{{\tt
  hep-ph/0603182}}].

\bibitem{Melnikov:2006kv}
K.~Melnikov and F.~Petriello, {\it {Electroweak gauge boson production at
  hadron colliders through $\mathcal{O}(\alpha_s^2)$}},  {\em Phys.Rev.} {\bf
  D74} (2006) 114017, [\href{http://xxx.lanl.gov/abs/hep-ph/0609070}{{\tt
  hep-ph/0609070}}].

\bibitem{Catani:2009sm}
S.~Catani, L.~Cieri, G.~Ferrera, D.~de~Florian, and M.~Grazzini, {\it {Vector
  boson production at hadron colliders: A Fully exclusive QCD calculation at
  NNLO}},  {\em PRLTA,103,082001.2009} {\bf 103} (2009) 082001,
  [\href{http://xxx.lanl.gov/abs/0903.2120}{{\tt arXiv:0903.2120}}].

\bibitem{Catani:2010en}
S.~Catani, G.~Ferrera, and M.~Grazzini, {\it {W boson production at hadron
  colliders: the lepton charge asymmetry in NNLO QCD}},  {\em JHEP} {\bf 1005}
  (2010) 006, [\href{http://xxx.lanl.gov/abs/1002.3115}{{\tt
  arXiv:1002.3115}}].

\bibitem{Alioli:2010qp}
S.~Alioli, P.~Nason, C.~Oleari, and E.~Re, {\it {Vector boson plus one jet
  production in POWHEG}},  {\em JHEP} {\bf 1101} (2011) 095,
  [\href{http://xxx.lanl.gov/abs/1009.5594}{{\tt arXiv:1009.5594}}].

\bibitem{Frixione:2007vw}
S.~Frixione, P.~Nason, and C.~Oleari, {\it {Matching NLO QCD computations with
  Parton Shower simulations: the POWHEG method}},  {\em JHEP} {\bf 0711} (2007)
  070, [\href{http://xxx.lanl.gov/abs/0709.2092}{{\tt arXiv:0709.2092}}].

\bibitem{Hamilton:2008pd}
K.~Hamilton, P.~Richardson, and J.~Tully, {\it {A Positive-Weight
  Next-to-Leading Order Monte Carlo Simulation of Drell-Yan Vector Boson
  Production}},  {\em JHEP} {\bf 0810} (2008) 015,
  [\href{http://xxx.lanl.gov/abs/0806.0290}{{\tt arXiv:0806.0290}}].

\bibitem{Frixione:2006gn}
S.~Frixione and B.~R. Webber, {\it {The MC@NLO 3.3 Event Generator}},
  \href{http://xxx.lanl.gov/abs/hep-ph/0612272}{{\tt hep-ph/0612272}}.

\bibitem{Baur:1997wa}
U.~Baur, S.~Keller, and W.~Sakumoto, {\it {QED radiative corrections to $Z$
  boson production and the forward backward asymmetry at hadron colliders}},
  {\em Phys.Rev.} {\bf D57} (1998) 199--215,
  [\href{http://xxx.lanl.gov/abs/hep-ph/9707301}{{\tt hep-ph/9707301}}].

\bibitem{Baur:1998kt}
U.~Baur, S.~Keller, and D.~Wackeroth, {\it {Electroweak radiative corrections
  to $W$ boson production in hadronic collisions}},  {\em Phys.Rev.} {\bf D59}
  (1999) 013002, [\href{http://xxx.lanl.gov/abs/hep-ph/9807417}{{\tt
  hep-ph/9807417}}].

\bibitem{Baur:2001ze}
U.~Baur, O.~Brein, W.~Hollik, C.~Schappacher, and D.~Wackeroth, {\it
  {Electroweak radiative corrections to neutral current Drell-Yan processes at
  hadron colliders}},  {\em Phys.Rev.} {\bf D65} (2002) 033007,
  [\href{http://xxx.lanl.gov/abs/hep-ph/0108274}{{\tt hep-ph/0108274}}].

\bibitem{Dittmaier:2001ay}
S.~Dittmaier and M.~Kr{\"a}mer, {\it {Electroweak radiative corrections to W
  boson production at hadron colliders}},  {\em Phys.Rev.} {\bf D65} (2002)
  073007, [\href{http://xxx.lanl.gov/abs/hep-ph/0109062}{{\tt
  hep-ph/0109062}}].

\bibitem{Dittmaier:2009cr}
S.~Dittmaier and M.~Huber, {\it {Radiative corrections to the neutral-current
  Drell-Yan process in the Standard Model and its minimal supersymmetric
  extension}},  {\em JHEP} {\bf 1001} (2010) 060,
  [\href{http://xxx.lanl.gov/abs/0911.2329}{{\tt arXiv:0911.2329}}].

\bibitem{Brensing:2007qm}
S.~Brensing, S.~Dittmaier, M.~Kr{\"a}mer, and A.~M{\"u}ck, {\it {Radiative
  corrections to $W^-$ boson hadroproduction: Higher-order electroweak and
  supersymmetric effects}},  {\em Phys.Rev.} {\bf D77} (2008) 073006,
  [\href{http://xxx.lanl.gov/abs/0710.3309}{{\tt arXiv:0710.3309}}].

\bibitem{Richardson:2010gz}
P.~Richardson, R.~Sadykov, A.~Sapronov, M.~Seymour, and P.~Skands, {\it {QCD
  parton showers and NLO EW corrections to Drell-Yan}},
  \href{http://xxx.lanl.gov/abs/1011.5444}{{\tt arXiv:1011.5444}}.

\bibitem{Zykunov:2008zz}
V.~Zykunov, {\it {Total calculation of electroweak corrections to the Drell-Yan
  process for LHC}},  {\em Phys.Atom.Nucl.} {\bf 71} (2008) 732--745.

\bibitem{Balossini:2008cs}
G.~Balossini, G.~Montagna, C.~Carloni~Calame, M.~Moretti, M.~Treccani, {\em
  et.~al.}, {\it {Electroweak and QCD corrections to Drell Yan processes}},
  {\em Acta Phys.Polon.} {\bf B39} (2008) 1675,
  [\href{http://xxx.lanl.gov/abs/0805.1129}{{\tt arXiv:0805.1129}}]. In honour
  of the 60th birthday of Prof. S. Jadach.

\bibitem{Balossini:2011zz}
G.~Balossini, C.~Carloni~Calame, G.~Montagna, M.~Moretti, O.~Nicrosini, {\em
  et.~al.}, {\it {Combining electroweak and QCD corrections to Drell-Yan
  processes at hadron colliders}},  {\em AIP Conf.Proc.} {\bf 1317} (2011)
  25--32.

\bibitem{Kilgore:2011pa}
W.~B. Kilgore and C.~Sturm, {\it {Two-Loop Virtual Corrections to Drell-Yan
  Production at order $\alpha_s \alpha^3$}},
  [\href{http://xxx.lanl.gov/abs/1107.4798}{{\tt arXiv:1107.4798}}].

\bibitem{Giele:1993dj}
W.~Giele, E.~Glover, and D.~A. Kosower, {\it {Higher order corrections to jet
  cross-sections in hadron colliders}},  {\em Nucl.Phys.} {\bf B403} (1993)
  633--670, [\href{http://xxx.lanl.gov/abs/hep-ph/9302225}{{\tt
  hep-ph/9302225}}].

\bibitem{vanderBij:1988ac}
J.~van~der Bij and E.~Glover, {\it {Z Boson production and decay via gluons}},
  {\em Nucl.Phys.} {\bf B313} (1989) 237.

\bibitem{Campbell:2002tg}
J.~M. Campbell and R.~Ellis, {\it {Next-to-leading order corrections to W+2 jet
  and Z+2 jet production at hadron colliders}},  {\em Phys.Rev.} {\bf D65}
  (2002) 113007, [\href{http://xxx.lanl.gov/abs/hep-ph/0202176}{{\tt
  hep-ph/0202176}}].

\bibitem{Hollik:2007sq}
W.~Hollik, T.~Kasprzik, and B.~Kniehl, {\it {Electroweak corrections to W-boson
  hadroproduction at finite transverse momentum}},  {\em Nucl.Phys.} {\bf B790}
  (2008) 138--159, [\href{http://xxx.lanl.gov/abs/0707.2553}{{\tt
  arXiv:0707.2553}}].

\bibitem{Kuhn:2007cv}
J.~H. K{\"u}hn, A.~Kulesza, S.~Pozzorini, and M.~Schulze, {\it {Electroweak
  corrections to hadronic production of W bosons at large transverse momenta}},
   {\em Nucl.Phys.} {\bf B797} (2008) 27--77,
  [\href{http://xxx.lanl.gov/abs/0708.0476}{{\tt arXiv:0708.0476}}].

\bibitem{Kuhn:2004em}
J.~H. K{\"u}hn, A.~Kulesza, S.~Pozzorini, and M.~Schulze, {\it {Logarithmic
  electroweak corrections to hadronic Z+1 jet production at large transverse
  momentum}},  {\em Phys.Lett.} {\bf B609} (2005) 277--285,
  [\href{http://xxx.lanl.gov/abs/hep-ph/0408308}{{\tt hep-ph/0408308}}].

\bibitem{Kuhn:2005az}
J.~H. K{\"u}hn, A.~Kulesza, S.~Pozzorini, and M.~Schulze, {\it {One-loop weak
  corrections to hadronic production of Z bosons at large transverse momenta}},
   {\em Nucl.Phys.} {\bf B727} (2005) 368--394,
  [\href{http://xxx.lanl.gov/abs/hep-ph/0507178}{{\tt hep-ph/0507178}}].

\bibitem{Kuhn:2007qc}
J.~H. K{\"u}hn, A.~Kulesza, S.~Pozzorini, and M.~Schulze, {\it {Electroweak
  corrections to large transverse momentum production of W bosons at the LHC}},
   {\em Phys.Lett.} {\bf B651} (2007) 160--165,
  [\href{http://xxx.lanl.gov/abs/hep-ph/0703283}{{\tt hep-ph/0703283}}].

\bibitem{Denner:2009gj}
A.~Denner, S.~Dittmaier, T.~Kasprzik, and A.~M{\"u}ck, {\it {Electroweak
  corrections to W + jet hadroproduction including leptonic W-boson decays}},
  {\em JHEP} {\bf 0908} (2009) 075,
  [\href{http://xxx.lanl.gov/abs/0906.1656}{{\tt arXiv:0906.1656}}].

\bibitem{Denner:2011vu}
A.~Denner, S.~Dittmaier, T.~Kasprzik, and A.~M{\"u}ck, {\it {Electroweak
  corrections to dilepton + jet production at hadron colliders}},  {\em JHEP}
  {\bf 1106} (2011) 069, [\href{http://xxx.lanl.gov/abs/1103.0914}{{\tt
  arXiv:1103.0914}}].

\bibitem{Ellis:2008qc}
R.~Ellis, W.~Giele, Z.~Kunszt, K.~Melnikov, and G.~Zanderighi, {\it {One-loop
  amplitudes for W+3 jet production in hadron collisions}},  {\em JHEP} {\bf
  0901} (2009) 012, [\href{http://xxx.lanl.gov/abs/0810.2762}{{\tt
  arXiv:0810.2762}}].

\bibitem{Berger:2009ep}
C.~Berger, Z.~Bern, L.~J. Dixon, F.~Febres~Cordero, D.~Forde, {\em et.~al.},
  {\it {Next-to-Leading Order QCD Predictions for W+3-Jet Distributions at
  Hadron Colliders}},  {\em Phys.Rev.} {\bf D80} (2009) 074036,
  [\href{http://xxx.lanl.gov/abs/0907.1984}{{\tt arXiv:0907.1984}}].

\bibitem{Berger:2010vm}
C.~Berger, Z.~Bern, L.~J. Dixon, F.~Febres~Cordero, D.~Forde, {\em et.~al.},
  {\it {Next-to-Leading Order QCD Predictions for Z,gamma+3-Jet Distributions
  at the Tevatron}},  {\em Phys.Rev.} {\bf D82} (2010) 074002,
  [\href{http://xxx.lanl.gov/abs/1004.1659}{{\tt arXiv:1004.1659}}].

\bibitem{Berger:2010zx}
C.~Berger, Z.~Bern, L.~J. Dixon, F.~Febres~Cordero, D.~Forde, {\em et.~al.},
  {\it {Precise Predictions for W + 4 Jet Production at the Large Hadron
  Collider}},  {\em Phys.Rev.Lett.} {\bf 106} (2011) 092001,
  [\href{http://xxx.lanl.gov/abs/1009.2338}{{\tt arXiv:1009.2338}}].

\bibitem{Gounaris:2007gx}
G.~Gounaris, J.~Layssac, and F.~M. Renard, {\it {Remarkable virtual SUSY
  effects in W+- production at high energy hadron colliders}},  {\em Phys.Rev.}
  {\bf D77} (2008) 013003, [\href{http://xxx.lanl.gov/abs/0709.1789}{{\tt
  arXiv:0709.1789}}].

\bibitem{Berge:2007dz}
S.~Berge, W.~Hollik, W.~M. M{\"o}sle, and D.~Wackeroth, {\it {SUSY QCD one-loop
  effects in (un)polarized top-pair production at hadron colliders}},  {\em
  Phys.Rev.} {\bf D76} (2007) 034016,
  [\href{http://xxx.lanl.gov/abs/hep-ph/0703016}{{\tt hep-ph/0703016}}].

\bibitem{Collins:1978wz}
J.~C. Collins, F.~Wilczek, and A.~Zee, {\it {Low-Energy Manifestations of Heavy
  Particles: Application to the Neutral Current}},  {\em Phys.Rev.} {\bf D18}
  (1978) 242.

\bibitem{Nason:1989zy}
P.~Nason, S.~Dawson, and R.~Ellis, {\it {The One Particle Inclusive
  Differential Cross-Section for Heavy Quark Production in Hadronic
  Collisions}},  {\em Nucl.Phys.} {\bf B327} (1989) 49--92.

\bibitem{Hahn:2000kx}
T.~Hahn, {\it {Generating Feynman diagrams and amplitudes with FeynArts 3}},
  {\em Comput.Phys.Commun.} {\bf 140} (2001) 418--431,
  [\href{http://xxx.lanl.gov/abs/hep-ph/0012260}{{\tt hep-ph/0012260}}].

\bibitem{Hahn:1998yk}
T.~Hahn and M.~Perez-Victoria, {\it {Automatized one loop calculations in
  four-dimensions and D-dimensions}},  {\em Comput.Phys.Commun.} {\bf 118}
  (1999) 153--165, [\href{http://xxx.lanl.gov/abs/hep-ph/9807565}{{\tt
  hep-ph/9807565}}].

\bibitem{Nogueira:1991ex}
P.~Nogueira, {\it {Automatic Feynman graph generation}},  {\em J.Comput.Phys.}
  {\bf 105} (1993) 279--289.

\bibitem{Form}
J.~Vermaseren, {\it {New features of FORM}},
  \href{http://xxx.lanl.gov/abs/math-ph/0010025}{{\tt math-ph/0010025}}.

\bibitem{Anastasiou:2004vj}
C.~Anastasiou and A.~Lazopoulos, {\it {Automatic integral reduction for higher
  order perturbative calculations}},  {\em JHEP} {\bf 0407} (2004) 046,
  [\href{http://xxx.lanl.gov/abs/hep-ph/0404258}{{\tt hep-ph/0404258}}].

\bibitem{Hahn:2004fe}
T.~Hahn, {\it {CUBA: A Library for multidimensional numerical integration}},
  {\em Comput.Phys.Commun.} {\bf 168} (2005) 78--95,
  [\href{http://xxx.lanl.gov/abs/hep-ph/0404043}{{\tt hep-ph/0404043}}].

\bibitem{Ellis:2007qk}
R.~Ellis and G.~Zanderighi, {\it {Scalar one-loop integrals for QCD}},  {\em
  JHEP} {\bf 0802} (2008) 002, [\href{http://xxx.lanl.gov/abs/0712.1851}{{\tt
  arXiv:0712.1851}}].

\bibitem{Jegerlehner:2000dz}
F.~Jegerlehner, {\it {Facts of life with gamma(5)}},  {\em Eur.Phys.J.} {\bf
  C18} (2001) 673--679, [\href{http://xxx.lanl.gov/abs/hep-th/0005255}{{\tt
  hep-th/0005255}}].

\bibitem{Chanowitz:1979zu}
M.~S. Chanowitz, M.~Furman, and I.~Hinchliffe, {\it {The Axial Current in
  Dimensional Regularization}},  {\em Nucl.Phys.} {\bf B159} (1979) 225.

\bibitem{Larin:1993tq}
S.~Larin, {\it {The Renormalization of the axial anomaly in dimensional
  regularization}},  {\em Phys.Lett.} {\bf B303} (1993) 113--118,
  [\href{http://xxx.lanl.gov/abs/hep-ph/9302240}{{\tt hep-ph/9302240}}].

\bibitem{Nakamura:2010zzi}
{\bf Particle Data Group} Collaboration, K.~Nakamura {\em et.~al.}, {\it
  {Review of particle physics}},  {\em J.Phys.} {\bf G37} (2010) 075021.

\bibitem{Martin:2009iq}
A.~Martin, W.~Stirling, R.~Thorne, and G.~Watt, {\it {Parton distributions for
  the LHC}},  {\em Eur.Phys.J.} {\bf C63} (2009) 189--285,
  [\href{http://xxx.lanl.gov/abs/0901.0002}{{\tt arXiv:0901.0002}}].

\bibitem{Whalley:2005nh}
M.~R. Whalley, D.~Bourilkov, and R.~C. Group, {\it {The Les Houches Accord PDFs
  (LHAPDF) and Lhaglue}},  \href{http://xxx.lanl.gov/abs/hep-ph/0508110}{{\tt
  hep-ph/0508110}}.

\bibitem{Collaboration:2011ida}
{\bf CMS} Collaboration, S.~Chatrchyan {\em et.~al.}, {\it {Search for New
  Physics with Jets and Missing Transverse Momentum in pp collisions at sqrt(s)
  = 7 TeV}},  \href{http://xxx.lanl.gov/abs/1106.4503}{{\tt arXiv:1106.4503}}.

\bibitem{Aad:2011xm}
{\bf ATLAS} Collaboration, G.~Aad {\em et.~al.}, {\it {Search for
  supersymmetric particles in events with lepton pairs and large missing
  transverse momentum in $\sqrt{s}=7$ TeV proton-proton collisions with the
  ATLAS experiment}},  \href{http://xxx.lanl.gov/abs/1103.6214}{{\tt
  arXiv:1103.6214}}.

\end{thebibliography}\endgroup
